\documentclass[aps,prd,superscriptaddress,amsmath,groupedaddress,twocolumn]{revtex4}

\usepackage{color,hangcaption,hhline,psfrag,rotating,amssymb}
\usepackage[hang,nooneline]{subfigure}
\usepackage{dcolumn}

\begin{document}
\title{Heavy meson masses and decay constants from relativistic heavy quarks in full lattice QCD}

\author{C. McNeile}
\affiliation{Bergische Universit\"{a}t Wuppertal, Gaussstr.\,20, D-42119 Wuppertal, Germany}
\author{C. T. H. Davies}
\email[]{c.davies@physics.gla.ac.uk}
\affiliation{SUPA, School of Physics and Astronomy, University of Glasgow, Glasgow, G12 8QQ, UK}
\author{E. Follana}
\affiliation{Departamento de F\'{\i}sica Te\'{o}rica, Universidad de Zaragoza, E-50009 Zaragoza, Spain}
\author{K. Hornbostel}
\affiliation{Southern Methodist University, Dallas, Texas 75275, USA}
\author{G. P. Lepage}
\affiliation{Laboratory of Elementary-Particle Physics, Cornell University, Ithaca, New York 14853, USA}

\collaboration{HPQCD collaboration}
\homepage{http://www.physics.gla.ac.uk/HPQCD}
\noaffiliation

\date{\today}

\begin{abstract}
We determine masses and decay constants of heavy-heavy and heavy-charm 
pseudoscalar mesons as a function of heavy quark mass
using a fully relativistic formalism known as Highly Improved 
Staggered Quarks for the heavy quark. 
We are able to cover the region from the charm quark 
mass to the bottom quark mass using MILC ensembles with lattice spacing 
values from 0.15 fm down to 0.044 fm. We obtain 
$f_{B_c}$ = 0.427(6) GeV; $m_{B_c}$ = 6.285(10) GeV 
and $f_{\eta_b}$ = 0.667(6) GeV. Our value for $f_{\eta_b}$ is 
within a few percent of $f_{\Upsilon}$ confirming that spin effects 
are surprisingly small for heavyonium decay constants. Our value 
for $f_{B_c}$ is significantly lower than potential model values 
being used to estimate production rates at the LHC. We
discuss the changing physical heavy-quark mass dependence of 
decay constants from heavy-heavy 
through heavy-charm to heavy-strange mesons. 
A comparison between the three 
different systems confirms that the $B_c$ system behaves in some 
ways more like a heavy-light system than a heavy-heavy one.  
Finally we summarise current results on decay constants of 
gold-plated mesons. 
\end{abstract}


\maketitle

\section{Introduction}
\label{sec:intro}

Lattice QCD calculations offer particular promise for 
$B$ meson physics where a number of relatively 
simple weak decay processes give access to elements of 
the CKM matrix that are important for constraining 
the unitarity triangle of the Standard Model~\cite{cdlat11}. The 
theoretical calculation of the appropriate weak 
matrix elements must be done with percent accuracy 
for stringent constraints, making optimal use of the experimental 
results. 
This has not yet been achieved, despite the enormous 
success of lattice QCD over the last five years 
and its acceptance as a precision
tool for QCD physics~\cite{ourlatqcd}. 
Work is ongoing on several different 
approaches. Here we continue discussion of an alternative 
method for $B$ meson physics that may offer a faster route to 
high accuracy for some quantities than other methods currently in use.  
Following work on accurate $b$ and $c$ quark masses~\cite{bcmasses} 
and heavy-strange 
decay constants~\cite{bshisq}, 
we show results for masses and decay constants of $B_c$ and $\eta_b$ 
mesons and map out their heavy-quark mass dependence. 
As well as showing that high accuracy can be achieved, these 
results provide an interesting comparison of how heavy-charm 
mesons sit between heavy-heavy and heavy-light. 

The calculations use a discretisation of the quark Lagrangian 
onto the lattice known as the Highly Improved Staggered Quark 
(HISQ) action~\cite{hisqdef}. This has the advantages of being numerically very 
fast along with having small discretisation errors and enough 
chiral symmetry (a PCAC relation) that the weak current that 
causes charged pseudoscalar mesons to decay leptonically 
is absolutely normalised.   
This action readily gives $\pi$ and $K$ meson decay constants 
with errors below 1\% on gluon field configurations 
that include the full effect 
of $u$, $d$ and $s$ quarks in the sea~\cite{fdsorig, dowdallr1}.
Results from multiple values of the lattice spacing and multiple 
sea $u/d$ quark masses allow extrapolation to the real world 
with physical 
$u/d$ quark masses at zero lattice spacing. 

The HISQ action gives similarly accurate results for mesons containing 
$c$ quarks~\cite{fdsorig}, 
significantly improving on previous methods
that use a nonrelativistic effective theory such as the 
Fermilab action~\cite{fermilab} or NRQCD~\cite{nrqcd}. 
The key advantages 
are clear: the HISQ action has no errors from 
missing higher order terms in the effective theory or 
from the renormalisation of 
the decay constant~\cite{cdlat11}. 
The price to be paid is that of the 
discretisation errors. These
errors are much larger for $c$ quarks than for $u/d$ and $s$, 
since their size is now set by $m_ca$ rather than $\Lambda_{QCD}a$. 
They can be well controlled, however, using the HISQ action on
gluon configurations with a wide range of 
lattice spacing values down to 0.045 fm where
$m_ca$ = 0.2~\cite{fdsupdate}.  
Discretisation errors are in fact the only issue for the 
$D_s$ meson, for which particularly accurate results can 
be obtained. 
This meson has no valence light quarks and 
the dependence of both its mass and decay constant on 
the $u/d$ quark masses is seen to be very small~\cite{fdsupdate}, 
meaning that uncertainties from the chiral extrapolation 
are not significant. 

It is less clear what to do for $b$ quarks because they 
are so much heavier. To achieve $m_ba <1$ we need a lattice spacing, 
$a <$ 0.04fm. Using NRQCD or the Fermilab formalism 
we can readily handle $b$ quarks 
on much coarser lattices, with $a \approx$ 0.1 fm, but must then 
take a substantial error (currently 
4\% for NRQCD~\cite{newfb}) 
from matching the weak annihilation current 
to full QCD perturbatively. Work is underway to reduce this error~\cite{jonna}. 
It should also be emphasised that this matching error is not present in 
ratios of decay constants, for example $f_{B_s}/f_B$ which is known 
to 2\% from NRQCD~\cite{newfb}. 

Here we show what accuracy is possible using 
the HISQ action for $b$ quarks. 
We use quark masses heavier than 
that of the $c$ quark and map out the heavy quark mass 
dependence of both masses and decay constants for a variety of different 
pseudoscalar mesons. 
By using experience from the $D_s$~\cite{fdsupdate} and 
concentrating on mesons that do not contain valence light quarks 
we do not have to worry significantly about the extrapolation to 
the physical $u/d$ quark mass limit. The key issue is that 
of discretisation errors, as for $f_{D_s}$, and we therefore 
work with the same large range of lattice spacing values from 
0.15fm to 0.044fm, so that we can account fully for the $a$ dependence.  
It is important to separate discretisation effects from physical dependence 
on the heavy quark mass since we do also have to extrapolate to the 
physical $b$ mass from the quark masses that we are able to reach 
on these lattices. We are only able to obtain results directly at close to 
the physical $b$ mass on the finest, 0.044fm lattice. 

We have already demonstrated how well this method works in 
determining the decay constant of the $B_s$ meson~\cite{bshisq}, one of 
the key quantities of interest for CKM studies. Mapping out 
the $B_s$ decay constant as a function of heavy quark mass 
showed that the decay constant peaks around the $D_s$ and 
then falls slowly. 
We found that $f_{B_s}/f_{D_s} = 0.906(14)$, 
the first significant demonstration that this ratio is 
less than 1. 
 
Here we extend this work to map out results for 
the decay constants of the $\eta_b$ 
and $B_c$ mesons, along with the $B_c$ meson mass. 
The $B_c$ meson mass is known experimentally but 
its leptonic decay rate has not yet been measured and 
so we provide the first prediction of that in full lattice QCD. 
The masses and decay constants also reveal information about 
the nature of these mesons that can provide useful input to model 
calculations. For example, does the $B_c$ meson 
look more like a heavy-heavy meson or a heavy-light meson?   
It is important to emphasise that both the results determined 
at the $b$ quark mass and the dependence on the heavy quark mass 
(and on any light quark masses) 
have physical meaning: the former can 
be tested against experiment but the latter can provide 
stringent tests of models and comparison between 
lattice QCD calculations. 

The layout of the paper is as follows: section~\ref{sec:latt} 
describes the lattice calculation and then section~\ref{sec:results} gives 
results for heavy-heavy and heavy-charm mesons in turn. We compare 
the $B_c$ meson mass to experiment and predict its decay constant 
as well as comparing the behaviour of 
heavy-charm mesons to that of heavyonium and 
heavy-strange mesons. 
Section~\ref{sec:conclusions} gives our conclusions,
looking forward to what will be possible for $b$ quark 
physics on even finer 
lattices in future. 

\section{Lattice calculation}
\label{sec:latt}

We use ensembles of lattice gluon 
configurations at 5 different, widely separated, values of 
the lattice spacing, provided by the MILC collaboration. 
The configurations include the effect of $u$, $d$ and $s$
quarks in the sea with the improved staggered (asqtad) formalism. 
Table~\ref{tab:params} lists the 
parameters of the ensembles. 
The $u$ and $d$ masses are taken to be the 
same, and the ensembles have $m_{u/d}/m_s$ approximately 0.2.
As discussed in section~\ref{sec:intro}, 
we expect sea quark mass effects to be small 
for the gold-plated mesons with no valence light
quarks that we study here. 

The lattice spacing is determined on an ensemble-by-ensemble basis 
using a parameter $r_1$ that comes from fits to the static quark 
potential calculated on the lattice~\cite{milcreview}. This 
parameter can be determined with very small statistical/fitting errors. 
However, its physical value is not accessible to experiment and 
so must be determined using other quantities, calculated on the 
lattice, that are. We have determined $r_1$ = 0.3133(23) fm
using four different quantities ranging from the (2S-1S) splitting 
in the $\Upsilon$ system to the decay constant of the $\eta_s$ (fixing 
$f_K$ and $f_{\pi}$ from experiment)~\cite{oldr1paper}. 
Using our value for $r_1$ and 
the MILC values for $r_1/a$ given in Table~\ref{tab:params} we can 
determine $a$ in fm on each ensemble or, equivalently, $a^{-1}$ in 
GeV needed to convert lattice masses to physical units. It is important 
to note that the relative values of $a$ (from $r_1/a$) are determined 
more accurately than the absolute values of $a$ (from $r_1$). Our fits 
account for this to give two separate errors in our error budgets. 

\begin{table}
\begin{tabular}{llllllll}
\hline
\hline
Set &  $r_1/a$ & $au_0m_{l}^{asq}$ & $au_0m_{s}^{asq}$ & $L/a$ & $T/a$ & $N_{conf}\times N_t$\\
\hline
1 & 2.152(5) & 0.0097 & 0.0484 & 16 & 48 & $631\times 2$ \\
\hline
2 &  2.618(3) & 0.01 & 0.05 & 20 & 64 & $595 \times 2$ \\
\hline
3 & 3.699(3) & 0.0062 & 0.031 & 28 & 96  & $566 \times 4$ \\
\hline 
4 &  5.296(7) & 0.0036 & 0.018 & 48 & 144 & $201 \times 2$ \\
\hline
5 & 7.115(20) & 0.0028 & 0.014 & 64 & 192 & $208 \times 2$ \\
\hline
\hline
\end{tabular}
\caption{Ensembles (sets) of MILC configurations used for this analysis. 
The sea 
asqtad quark masses $m_l^{asq}$ ($l = u/d$) and $m_s^{asq}$ 
are given in the MILC convention where $u_0$ is the plaquette 
tadpole parameter. 
The lattice spacing values in units of $r_1$ after `smoothing'
are given in the second column~\cite{milcreview}. 
Set 1 is `very coarse'; set 2, `coarse'; set 3, 
`fine'; set 4 `superfine' and set 5 `ultrafine'.  The size of 
the lattices is given by $L^3 \times T$. The final column gives the 
number of configurations used and the number of time sources for propagators 
per configuration. }
\label{tab:params}
\end{table}

Table~\ref{tab:params} lists the number 
of configurations used from each ensemble and the number of time sources 
for the valence HISQ propagators per configuration. 
To increase statistics further we use a `random wall' 
source for the quark propagators from a given time source. 
When quark propagators are combined this 
effectively increases the number of meson correlators 
sampled and reduces the statistical noise by a large factor for the case 
of pseudoscalar mesons. 
We also take a random starting point for our time sources 
for the very coarse, coarse and fine 
ensembles. 

We use many different masses for the HISQ 
valence quarks varying 
from masses close to that of the $s$ quark 
to much heavier values for $c$ quarks and for quarks with 
masses between $c$ and $b$.  On all sets the largest valence 
quark mass in lattice units that we use is $m_ha = 0.85$. 
These propagators are combined to make goldstone pseudoscalar 
meson correlators at zero momentum 
with all possible combinations of valence quark masses. 
We separate them into `heavy-heavy' correlators when both masses 
are the same and are close to charm or heavier; `heavy-charm' when 
one mass is close to charm and the other is heavier and `heavy-strange' 
when one mass is close to strange and the other is close to charm 
or heavier. 

The meson correlation function is averaged over time sources 
on a single configuration so that any correlations between 
the time sources are removed. 
Autocorrelations between results on 
successive configurations in an ensemble were visible by binning only 
on the finest lattices. 
We therefore bin the 
correlators on superfine and ultrafine lattices by a factor of two. 

The meson correlators are fit as a function of the 
time separation between source and sink, $t$, to the form: 
\begin{equation}
\overline{C}(t) = \sum_i a_i (e^{-M_i t} + e^{-M_i (T-t)})
\label{eq:fit1}
\end{equation}
for the case of equal mass quark and antiquark. 
$i=1$ is the ground state and larger $i$ values 
denote radial or other excitations with the same $J^{PC}$ quantum 
numbers. $T$ is the time extent of the lattice. 
For the unequal mass case there are additional 
`oscillating' terms coming from opposite parity states, denoted $i_p$: 
\begin{equation}
\overline{C}(t) = \sum_{i,i_p} a_i e^{-M_i t} + (-1)^ta_{i_p}e^{-M_{i_p} t} + ( t \rightarrow T- t)
\label{eq:fit2}
\end{equation}

To fit we use a number of exponentials $i$, and where appropriate $i_p$, 
in the range 2--6, loosely constraining the higher order exponentials 
by the use of Bayesian priors~\cite{gplbayes}. 
As the number of exponentials increases, 
we see the $\chi^2$ value fall below 1 and the results
for the fitted values 
and their errors for the parameters for the ground state $i=1$ stabilise. 
This allows us to determine the ground state parameters $a_1$ and $M_1$ as accurately 
as possible whilst allowing the full systematic error from the presence 
of higher excitations in the correlation function. 
We take the fit parameters 
to be the logarithm of the ground state masses $M_1$ and $M_{1_p}$ 
and the logarithms of 
the differences in mass between successive radial excitations (which 
are then forced to be positive). 
The Bayesian prior value for $M_1$ is obtained from a simple `effective 
mass' in the correlator and the prior width on the value is taken as a 
factor of 1.5. 
The prior value for the mass splitting between higher excitations 
is taken as roughly 600 MeV with a width of 300 MeV. 
Where oscillating states appear in the fit, 
the prior value for $M_{1_p}$
is taken as roughly 600 MeV above $M_1$ with a 
prior width of 300 MeV and the 
splitting between higher oscillating excitations is taken to be 
the same as for the non-oscillating states. 
The amplitudes $a_i$ and $a_{i_p}$ are given prior 
widths of 1.0. 
We apply a cut on the range of eigenvalues from the 
correlation matrix that are used in the fit of $10^{-3}$ 
or $10^{-4}$. We also cut out very small $t$ values from our 
fit, 
typically below 3 or 4, to reduce the effect of higher 
excitations. 

The amplitude, $a_1$, from the fits in 
equations~(\ref{eq:fit1}) and~(\ref{eq:fit2})
is directly related to the matrix element for the local 
pseudoscalar operator to create or destroy the ground-state pseudoscalar meson 
from the vacuum. Using the PCAC relation this can be related 
to the matrix element for the temporal axial current and thence to 
the decay constant. The PCAC relation guarantees that no 
renormalisation of the decay constant is needed. We have:  
\begin{equation}
f_P = (m_a + m_b) \sqrt{\frac{2a_1}{M_1^3}}.
\label{eq:atof}
\end{equation}
for meson $P$. 
Here $m_a$ and $m_b$ are the quark masses used in 
the lattice QCD calculation. 

$f_P$ is clearly a measure of the internal structure 
of a meson and in turn is related, 
for charged pseudoscalars such as the $\pi$, $K$, 
$D$, $D_s$, $B$ and $B_c$ mesons, 
to the experimentally measurable leptonic branching fraction via a $W$ boson:
\begin{equation}
{\cal{B}}(P \rightarrow l \nu_l (\gamma)) = \frac{G_F^2 |V_{ab}|^2\tau_P}{8\pi}f_{P}^2m_l^2m_{P}\left( 1-\frac{m_l^2}{m_{P}^2}\right)^2,
\label{eq:gamma}
\end{equation}
up to calculable electromagnetic corrections. 
$V_{ab}$ is the appropriate CKM element for quark 
content $a\overline{b}$. $\tau_P$ is the pseudoscalar meson lifetime. 
For neutral mesons there is no possibility to annihilate to a single 
particle via the temporal axial current. 
However, in the Standard Model the $B_s$ and $B$ are expected to annihilate to 
$\mu^+\mu^-$ with a rate that is proportional 
to $f_{P}^2|V_{tb}^*V_{tq}|^2$ via 4-fermion operators 
in the effective weak Hamiltonian~\cite{buras}. 
For the heavy-heavy pseudoscalar, the decay rate to two 
photons is related to its decay constant but 
only at leading order in a nonrelativistic expansion. 
In section~\ref{sec:results} we compare the pseudoscalar 
decay constant to that of its associated vector meson, 
determined directly from its decay to leptons. 

The results for masses and decay constants from 
fits in eqs.~(\ref{eq:fit1}) 
and~(\ref{eq:fit2}) and using eq.~(\ref{eq:atof})
are in units of the lattice spacing, and given in this form in 
the tables of section~\ref{sec:results}. 
To convert to physical units, as discussed earlier, 
we determine the lattice spacing using the parameter $r_1$. 

We then fit the results in physical units as a 
function of heavy quark mass to determine the heavy 
quark mass dependence and the physical value at 
the $b$ quark mass. Because the bare heavy quark 
mass used in the lattice action runs with lattice spacing
we need a proxy for it that is a physical 
quantity, such as a meson mass. In~\cite{bshisq} 
we used the heavy-strange pseudoscalar mass since 
we were focussing on heavy-strange mesons. Here we 
choose the mass of the heavy-heavy pseudoscalar 
meson, $\eta_h$, to provide the same $x$-axis 
for all of our plots showing dependence on the heavy quark 
mass. The positions of $c$ and $b$  
on these plots are then determined by the 
values of the $\eta_c$ and $\eta_b$ masses.

The experimental results for the $\eta_b$ and $\eta_c$ meson 
masses are 9.391(3) GeV and 2.981(1) GeV respectively~\cite{pdg}. 
Our lattice QCD calculation, however, is missing some ingredients 
from the real world which means that we must adjust the 
experimental values we use in our calibration. 
The key missing ingredients are electromagnetism, $c$ quarks 
in the sea and the possibility for the $\eta_b$ and $\eta_c$ mesons 
to annihilate to gluons, which we do not allow for in 
determining our $\eta_c$ and $\eta_b$ correlators. 
These effects all act in the same direction, that of lowering 
the meson mass in the real world compared to that in our lattice 
QCD world. 
We estimate the total shift from these effects for the 
$\eta_c$ to be -5.4(2.7) MeV and for the $\eta_b$, 
as -9(6) MeV~\cite{gregory}. 
The appropriate `experimental' masses for the $\eta_c$ and $\eta_b$ 
for our calculations are then 2.986(3) GeV and 9.400(7) GeV. 

\begin{table}
\begin{tabular}{lccc}
\hline
\hline
 &  electromagnetism & $c$-in-sea & annihiln to g\\
\hline
$M_{\eta_b}$ & -1.6(8) & -5(3) & -2.4(2.4)\\
$M_{\eta_c}$ & -2.6(1.3) & -0.4(2) & -2.4(1.2)\\
$M_{B_c}$ & +2(1) & -1(1) & -\\
$M_{B_s}$ & -0.1(1) & - & -\\
$M_{D_s}$ & +1.3(7) & - & -\\
\hline
\hline
\end{tabular}
\caption{Estimates of shifts in MeV to be applied to the 
masses determined in lattice QCD to allow for missing
electromagnetism, $c$ quarks in the sea and annihilation 
to gluons for the $\eta_b$ and $\eta_c$ mesons~\cite{gregory, fdsupdate}. 
The electromagnetic shift is estimated from a potential 
model for $\eta_b$, $\eta_c$ and $B_c$ and from a comparison 
of charged and neutral meson masses for $B_s$ and $D_s$. 
The $c$-in-sea and gluon annihilation shifts are estimated 
from perturbation theory. 
The errors on the shifts are given in brackets. 
Note that for the $\eta_S$ there are no shifts because 
the mass is fixed in lattice QCD~\cite{oldr1paper}. 
}
\label{tab:adjust}
\end{table}

Since we need to allow for the three ingredients
missing from our lattice QCD calculation when 
we determine meson masses in section~\ref{sec:results}, 
we give in Table~\ref{tab:adjust} a summary of our 
estimates of these effects~\cite{gregory, fdsupdate}. 
These estimates will be used to shift the lattice 
QCD results for comparison to experiment. 
Effects on decay constants are much smaller and we 
do not apply shifts in that case but simply include 
an additional uncertainty in the error budget.  

\begin{table}
\begin{tabular}{lccccc}
\hline
\hline
 &  form of $f_0$  & $b$ & $A$ & $c_{0000}$ & $c_{ijkl}$ \\
\hline
$f_{\eta_h}$ & $A({M}/{M_0})^b$ & $0\pm 1$ & $0 \pm 2$ & 1 & $0 \pm 4.5$\\
$\Delta_{H_s,hh}$ & $A({M}/{M_0})^b$ & 1 & 1 & $0 \pm 2$ & $0 \pm 1.5$ \\ 
$f_{H_s}$ & $A(\frac{\alpha_V(M)}{\alpha_V(M_{\eta_c})})^{-2/9}(\frac{M}{M_0})^b$ & -0.5 & $0 \pm 2$ & 1 & $0 \pm 1.5$ \\
$\Delta_{H_c,hh}$ & $A({(M-M_{\eta_c})}/{M_0})^b$ & 1 & 1 & $0 \pm 2$ & $0 \pm 1.5$ \\
$f_{H_c}$ & $A(\frac{\alpha_V(M)}{\alpha_V(M_{\eta_c})})^{-2/9}(\frac{M}{M_0})^b$ & -0.5 & $0 \pm 2$ & 1 & $0 \pm 3$ \\
$\Delta_{H_c,hs}$ & $A({M}/{M_0})^b$ & 0 & 1 & $0 \pm 2$ & $0 \pm 1.5$ \\
\hline
\hline
\end{tabular}
\caption{The functional form for $f_0(M)$, the 
leading power dependence on the heavy quark 
mass, used in fitting the different quantities 
described in section~\ref{sec:results} using 
equation~(\ref{eq:fitform}). The third and fourth columns 
give the prior values and widths for the parameters $A$ and $b$. 
In most cases $b$ was fixed and then a single number 
is given. Likewise the sixth column gives the 
prior value and width for the $c_{ijkl}$ where 
the sum was normalised so that $c_{0000}$ was set 
equal to 1.  
}
\label{tab:f0}
\end{table}

For consistency we fit a similar functional form 
to all quantities. This form must take account of 
physical heavy quark mass-dependence; discretisation 
errors and, for heavy-charm and heavy-strange 
mesons, mistuning of $c$ and $s$ quark masses.
We use the standard constrained fitting techniques 
that we earlier applied to the correlators~\cite{gplbayes}. 
For the dependence on heavy quark mass and 
lattice spacing for each set of results $f(M,a)$, we use 
\begin{eqnarray}
\label{eq:fitform}
&& f(M,a) = f_0(M) \times \nonumber \\ 
&&\sum_{i=0}^7 \sum_{j,k=0}^3 \sum_{l=0}^1 c_{ijkl} \left(\frac{M_0}{M}\right)^i (\frac{am_1}{\pi})^{2j}(\frac{am_2}{\pi})^{2k} (\frac{a\Lambda}{\pi})^{2l} \nonumber \\
&& + \delta f_s + \delta f_c 
\end{eqnarray}
The quantity that we use for the heavy 
quark mass, $M$, is given by $M = M_{\eta_h}$. 
$f_0$ is a function giving the `leading power'
behaviour expected for each quantity. This is 
either derived from HQET or potential model 
expectations and takes the general form 
$A(M/M_0)^b$. For the decay constants 
$f_{H_c}$ and $f_{H_s}$ we multiply this by the 
ratio of $\alpha_s$ values at the $b$ and the $c$ 
raised to the power of $-2/\beta_0$ = -2/9 for $n_f=3$. 
This is the expected prefactor from resumming leading 
logarithms in HQET~\cite{neubert}. For $\alpha_s$ we take 
$\alpha_V$ from lattice QCD~\cite{bcmasses, alpha}. 
We take $M_0$ to have the 
value 2 GeV so that the factor $M_0/M$ is approximately 
$1 \mathrm{GeV}/m_b$. We tabulate the different forms 
for $f_0$ in Table~\ref{tab:f0} along with 
the prior values taken for $A$ and $b$. $b$ is 
allowed to float for the fit to $f_{\eta_h}$. In other 
cases it is fixed to the expected value but we have 
checked that allowing it to float returns the 
expected value within errors. We take the 
same prior for $A$ of $0 \pm 2$ in all cases.  

The sum to the right of the leading term 
includes higher order corrections to the 
physical mass-dependence. These take the form 
of powers of $M_0/M$, again using $M_0 = 2\, \mathrm{GeV}$. 
We allow for 8 terms in the sum so that there 
is enough leeway to describe (by Taylor's theorem) 
any physically reasonable 
functional form in the fixed mass range from $c$ to 
$b$. For the heavy-charm case we in fact fit from 
$M = 4 \, \mathrm{GeV}$ upwards so that the functional form 
is that appropriate to the unequal valence mass case.  

The other terms in the sum of eq.~(\ref{eq:fitform}) allow for 
systematic errors resulting from 
sensitivity to the lattice spacing. Such 
discretisation errors depend 
on the lattice momentum cut-off, $\pi/a$, but  
can have a scale set by the different masses 
involved in the quantity under study. We allow 
for discretisation errors appearing with a 
scale of $m_1$ and $m_2$, where $m_1$ and 
$m_2$ are the two quark masses in the meson 
(they will be the same in heavyonium). 
To be conservative we allow in addition further 
discretisation errors with a scale of $\Lambda_{\rm QCD}$ 
where we take $\Lambda_{\rm QCD}$ = 0.5 GeV.  
The powers of lattice spacing that appear in 
the terms must be even since discretisation errors 
only appear as even powers for staggered quarks. 
For the decay constants the $c_{ijkl}$ are 
normalised so that $c_{0000}=1$.  
For the mass differences the fits are normalised 
so that $A$ is 1 and $c_{0000}$ floats. This is 
simply so that the fit can allow for significant 
discretisation errors when the 
physical mass difference is very small (particularly for the 
case of the $B_c$ to be discussed in section~\ref{sec:heavycharm}). 
The prior values for the other 
$c_{ijkl}$ are taken to be the same 
for all $i$, $j$, $k$ and $l$ but vary depending on the size of 
discretisation errors for the quantity being fit. 
They are larger for heavyonium than for heavy-strange 
quantities, for example. 
The values used are tabulated in Table~\ref{tab:f0}. 

The mistuning of the strange and charm quark 
masses, where relevant, can be handled very simply because our 
tuning of these masses is in fact very good. 
We simply include an additional additive factor in the fit of 
\begin{eqnarray}
\delta f_s &=& (c_s+\frac{d_s}{M} +e_s\left[\left(\frac{am_1}{\pi}\right)^2 + \left(\frac{am_2}{\pi}\right)^2 \right])\times \nonumber \\
&&(m_{\eta_s,\mathrm{latt}}^2 - m_{\eta_s,\mathrm{contnm}}^2)
\label{eq:mstune}
\end{eqnarray}
for heavy-strange mesons and 
\begin{eqnarray}
\delta f_c &=& (c_c+\frac{d_c}{M} + e_c\left[\left(\frac{am_1}{\pi}\right)^2 + \left(\frac{am_2}{\pi}\right)^2 \right])\times \nonumber \\ 
&&(m_{\eta_c,\mathrm{latt}} - m_{\eta_c,\mathrm{contnm}})
\label{eq:mctune}
\end{eqnarray}
for heavy-charm mesons. 
The forms above allow for linear quark mass dependence away 
from the tuned point. We do not need to include higher 
order terms because we are so close to the tuned point but 
we do allow for an $M$-dependent slope with discretisation 
errors (although in most 
cases neither of these additions makes any difference). 

To tune the strange quark mass we use
the $\eta_s$, an unphysical $s\overline{s}$ pseudoscalar 
meson whose valence quarks are not allowed to annihilate. 
Lattice QCD simulations show that its mass 
$m_{\eta_s, \mathrm{contnm}} = 0.6858(40) 
\mathrm{GeV}$~\cite{oldr1paper} when the strange quark mass is tuned 
(from the $K$ meson). Being a light pseudoscalar meson, the square 
of its mass is proportional to the quark mass. 
To tune the $c$ quark mass we use the $\eta_c$ meson, 
as discussed earlier. The $\eta_c$ meson is far from the 
light quark limit and so the meson mass is simply 
proportional to the quark mass. 
$c_s$ and $c_c$ are dimensionful coefficients that represent 
physical light quark mass dependence and can be compared 
between lattice QCD calculations and with models. 

We do not include correlations between the results 
for different $M$ on a given ensemble. We have not measured 
these correlations and the empirical Bayes criterion suggests 
that they are small. If we include a correlation matrix by hand
for the results it makes very little difference, a fraction 
of a standard deviation, to the final 
results. 

We also do not include effects from sea quark mass-dependence 
but, based on earlier work~\cite{fdsupdate}, we are 
able to estimate an uncertainty for that in our final 
results. 

\section{Results }
\label{sec:results}

\subsection{$f_{\eta_b}$}
\label{sec:etab}

The correlators for pseudoscalar heavyonium mesons 
have very little noise and we can readily 
obtain ground-state masses with statistical errors in the fourth or fifth 
decimal place and ground-state decay constants with errors of 0.1\%. 
Our results on each ensemble are given in Table~\ref{tab:etahresults}. 

\begin{table}
\begin{tabular}{lllll}
\hline
\hline
Set & $m_ha$ & $\epsilon$ & $M_{\eta_h}a$ & $f_{\eta_h}a$  \\
\hline
1 & 0.66 & -0.244 & 1.92020(16) & 0.3044(4)  \\
  & 0.81 & -0.335 & 2.19381(16) & 0.3491(5) \\
  & 0.825 &-0.344 & 2.22013(15) & 0.3539(5) \\
  & 0.85 & -0.359 & 2.26352(15) & 0.3622(5) \\
\hline
2 & 0.44 & -0.12 & 1.42402(13) & 0.21786(21)  \\ 
  & 0.63 & -0.226 & 1.80849(11) & 0.25998(20) \\
  & 0.66 & -0.244 & 1.86666(10) & 0.26721(20)  \\
  & 0.72 & -0.28 & 1.98109(10) & 0.28228(22)  \\
  & 0.753 & -0.3 & 2.04293(10) & 0.29114(24) \\ 
  & 0.85 & -0.36 & 2.21935(10) & 0.31900(27) \\
\hline
3 & 0.3 & -0.06 & 1.03141(8) & 0.15205(11)  \\
  & 0.413 & -0.107 & 1.28057(7) & 0.17217(11) \\
  & 0.43 & -0.115 & 1.31691(7) & 0.17508(11)  \\
  & 0.44 & -0.12 & 1.33816(7) & 0.17678(11)  \\
  & 0.45 & -0.125 & 1.35934(7) & 0.17850(11)  \\
  & 0.7 & -0.27 & 1.86536(5) & 0.22339(12) \\
  & 0.85 & -0.36 & 2.14981(5) & 0.25658(12) \\
\hline 
4 & 0.273 &  -0.0487 & 0.89935(10) & 0.11864(24)  \\
  & 0.28 & -0.051 & 0.91543(8) & 0.11986(21) \\
  & 0.564 & -0.187 & 1.52542(6) & 0.16004(16) \\
  & 0.705 & -0.271 & 1.80845(6) & 0.18071(16) \\
  & 0.76 & -0.305 & 1.91567(6) & 0.18962(17) \\
  & 0.85 & -0.359 & 2.08753(6) & 0.20576(16) \\
\hline
5 & 0.193 & -0.0247 & 0.66628(13) & 0.0882(3) \\
 & 0.195 & -0.02525 & 0.67117(6) & 0.08846(11) \\
  & 0.4 &  -0.101 & 1.13276(7) & 0.1149(4) \\
  & 0.5  & -0.151 & 1.34477(8) & 0.1260(5) \\
  & 0.7 & -0.268 & 1.75189(7) & 0.1498(5) \\
  & 0.85 & -0.359 & 2.04296(7) & 0.1708(6) \\
\hline
\hline
\end{tabular}
\caption{ Results for the masses and decay constants in lattice units 
of the goldstone pseudoscalars made 
from valence HISQ heavy quarks on the different 
MILC ensembles, enumerated in Table~\ref{tab:params}. Columns 2 and 3 give 
the corresponding bare heavy quark mass and the $\epsilon$ parameter, 
calculated at tree-level in $m_ha$~\cite{fdsupdate}. 
This corresponds to a coefficient for the Naik 3-link discretisation 
correction of $1+\epsilon$. 
Meson masses from fitting these correlators using a simpler 
fitting form are given in~\cite{bcmasses}. Results given here 
are in agreement but somewhat more accurate.  
The results for heavy quark masses close to charm 
are also given in~\cite{fdsupdate}. 
 }
\label{tab:etahresults}
\end{table}

\begin{figure}
\begin{center}
\includegraphics[width=0.9\hsize]{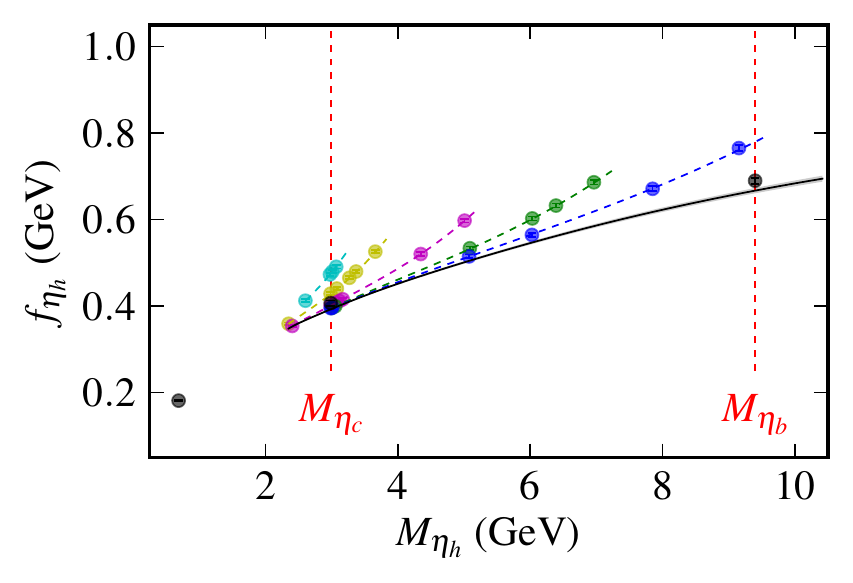}
\end{center}
\caption{Results for the pseudoscalar heavyonium decay constant 
plotted as a function of the pseudoscalar heavyonium mass. Results 
for very coarse, 
coarse, fine, superfine and ultrafine lattices appear from left 
to right. The colored dashed lines give the fitted 
function for that 
lattice spacing. The black line with grey error band gives 
the physical curve derived from our fit. 
The black circles with error bars at $M_{\eta_c}$ and $M_{\eta_b}$ 
are the values for the heavyonium vector decay constant 
at these physical points derived from the experimental 
leptonic widths for the $J/\psi$ and $\Upsilon$. 
The left-most black circle 
corresponds to the fictitious pseudoscalar $\eta_s$ particle 
whose decay constant was determined in~\cite{oldr1paper}.}
\label{fig:fhh}
\end{figure}

Results for $f_{\eta_h}$ are plotted against 
$M_{\eta_h}$ in Figure~\ref{fig:fhh}. 
Discretisation errors are apparent 
in this plot and lead to results at each 
value of the lattice spacing deviating 
substantially from the physical curve 
as the quark mass is increased. 
We fit the results to a 
physical curve allowing for discretisation 
errors as a function of the mass, 
as described in section~\ref{sec:latt} and using 
the priors from Table~\ref{tab:f0}. 
The power, $b$, in eq.~(\ref{eq:fitform}) 
is allowed to float in the fit. 

The obvious approach from which to gain 
some physical insight in this case is 
that of the nonrelativistic potential model. 
In its simplest form this involves 
solving Schr\"{o}dinger's equation 
for the wavefunction of a two-particle system with 
reduced mass $\mu$ ($= m_b/2$ for two 
$b$ quarks) in a potential $V(r)$ which is a function 
of the radial separation, $r$. 
At short distances we expect a Coulomb-like 
potential from QCD, and at large distances 
a string-like linear potential. However, other 
phenomenological forms that interpolate 
between these two  at intermediate distances 
also work well at reproducing the bound-state 
spectrum, see for example~\cite{kwongrosner}. The wavefunction 
is useful for a first approximation in calculations 
of transition rates. In this sense,
 the wavefunction at the origin, $\psi(0)$, can be 
related to the decay constant 
by $\psi(0) = f_{\eta_h}\sqrt{M_{\eta_h}/12}$.
However, $\psi(0)$ must be renormalised before 
it can be related to a physical matrix element 
and some of the radiative corrections are 
very substantial~\cite{kwongrosner}. In addition 
values of $\psi(0)$ vary widely with different 
forms for the potential that reproduce the same 
bound state spectrum because the spectrum itself 
provides little constraint on the potential at 
short distances~\cite{eichtenquigg}. Here we will 
make comparisons of our lattice QCD results to 
those from potential models but it is important 
to realise that the lattice QCD results for 
decay constants represent well-defined matrix elements 
in QCD and not model calculations. 

For a potential model with potential 
$r^N$ power counting arguments 
yield $\psi(0) \propto \mu^{3/(4+2N)}$ 
(see, for example,~\cite{close}). 
Then we would expect our fit for $f_{\eta_h}$ 
to need $b=1$ for $N=-1$ 
but $b=0$ for $N=1$, the two extremes 
of the QCD heavy quark potential. 
Simply from comparing values at $c$ and 
$b$ we might infer $b \approx 0.5$. 
In fact our fit gives 
the result $b = -0.08(10)$ but with significant 
power corrections in $1/M$, so that a simple 
power in $M$ does not describe the results 
using our parameterisation. 
The physical curve that we extract of dependence on 
the heavyonium meson mass is shown as 
the grey band in Figure~\ref{fig:fhh}.  

The fit has $\chi^2$ of 1.2 for 29 degrees of 
freedom and allows us to extract results 
for $c$ and $b$ quarks. The result for 
$f_{\eta_c}$ agrees within $1 \sigma$ with our earlier 
result of 0.3947(22) GeV~\cite{fdsupdate} 
where we fit results at $c$ only but included 
additional ensembles at different values of the 
sea $u/d$ quark masses. 
Results for $b$ quarks give:
\begin{eqnarray}
f_{\eta_b} &=& 0.667(6)(2) \mathrm{GeV} \nonumber \\
f_{\eta_b}/f_{\eta_c} &=& 1.698(13)(5) .
\label{eq:fetahres}
\end{eqnarray}
The first error comes from the fit and the second 
from additional systematic errors from effects 
not included in our lattice QCD calculation, i.e. 
electromagnetism, $c$ quarks in the sea and 
(since we have not extrapolated to physical 
$u/d$ sea quark masses here) sea quark mass 
effects. Both errors are split into their 
component parts in the error budget 
of Table~\ref{tab:etaherrors}.
We estimated the effects of electromagnetism on $f_{\eta_c}$ from 
a potential model in~\cite{fdsupdate}. We take the same 
0.4\% error for $f_{\eta_b}$ since it is a more tightly 
bound particle but with smaller electromagnetic charges. 
There is then some cancellation of the effect in the 
ratio $f_{\eta_b}/f_{\eta_c}$. 
The effects of $c$ quarks in the sea were shown to 
be similar to that of the hyperfine potential in~\cite{fdsupdate}
and the effect on $f_{\eta_h}$ can then be estimated
from the difference between $f_{\eta_h}$ and its associated 
vector particle. This is very small as we show below. 
We therefore expect that missing $c$ in the sea 
has a negligible effect
on $f_{\eta_c}$ and we estimate 0.2\% on $f_{\eta_b}$ where 
it is magnified by $(m_b/m_c)^2$. 
Sea quark mass effects on $f_{\eta_c}$ were shown to 
be very small in~\cite{fdsupdate}, at the same level as the 
statistical errors of 0.1\%. For $f_{\eta_b}$ we expect 
even smaller effects because it is a smaller particle. 
We take a 0.1\% error nevertheless, but allow for 
some cancellation in the ratio of $f_{\eta_b}/f_{\eta_c}$.   

\begin{table}
\begin{tabular}{lcc}
\hline
\hline
Error & $f_{\eta_b}$ & $f_{\eta_b}/f_{\eta_c}$ \\
\hline
statistics  & 0.6 &  0.6 \\
$M$ extrapoln  & 0.2  & 0.1 \\
$a^2$ extrapoln  & 0.5 & 0.4 \\
$r_1$  & 0.4 &  0.1 \\
$r_1/a$  & 0.5 &  0.3 \\
$M_{\eta_{c}}$ & 0.00 & 0.05 \\
sea quark mass effects & 0.1 & 0.05 \\
electromagnetism & 0.4 & 0.2 \\
$c$ in the sea & 0.2 & 0.2 \\
\hline
Total (\%) & 1.0 & 0.9 \\
\hline
\hline
\end{tabular}
\caption{ Full error budget for $f_{\eta_b}$ and the ratio 
$f_{\eta_b}/f_{\eta_c}$ in \%. See text for a fuller 
description of each error. The total error is 
obtained by adding the individual errors in quadrature.  
 }
\label{tab:etaherrors}
\end{table}

The two rightmost black points (at $M_{\eta_b}$ and 
$M_{\eta_c}$) in Fig.~\ref{fig:fhh} 
give the experimental values for the decay 
constants of the corresponding vector heavyonium mesons, $J/\psi$ 
and $\Upsilon$, for comparison to the results calculated 
here in lattice QCD for the $\eta_c$ and $\eta_b$. 
The decay constant for a vector meson can be defined by: 
\begin{equation}
\sum_i <0|\overline{\psi}\gamma_i\psi | V_i>/3 = f_V m_V.
\label{eq:fvec}
\end{equation}
It has the advantage here that it can be extracted 
very accurately from experiment because vector 
heavyonium mesons can annihilate, 
through the vector currrent, to a photon, 
seen as two leptons in the final state. 
The relationship 
between the leptonic decay width and the decay constant 
is: 
\begin{equation}
\Gamma(V_h \rightarrow e^+e^-) = \frac{4\pi}{3}\alpha_{QED}^2 e_h^2 \frac{f_V^2}{m_V}
\label{eq:vdecay}
\end{equation} 
where $e_h$ is the electric charge of the heavy quark 
in units of $e$. 
The experimental results~\cite{pdg} 
give $f_{J/\psi}$ = 407(5) MeV 
and $f_{\Upsilon}$ = 689(5) GeV, remembering 
that the electromagnetic coupling constant runs with 
scale and using $1/\alpha_{QED}(m_c) = 134$ and 
$1/\alpha_{QED}(m_b) = 132$\cite{alpha-em}. Thus 1\% accurate 
results for this decay constant are available 
from experiment, and can be used to test 
lattice QCD. Lattice QCD calculations of the 
$\Upsilon$ decay constant can be done~\cite{gray} but they are not 
yet as accurate as the results we give here for 
the $\eta_b$.  

The surprising result that we find on comparing 
the vector decay constant from experiment to the 
pseudoscalar decay constant from lattice QCD is 
how close they are. In the nonrelativistic limit, 
where spin effects disappear, the vector and pseudoscalar 
become the same particle. 
Away from this point, however, there can be substantial 
relativistic corrections, particularly for charmonium. 
Instead we find that the pseudoscalar decay constant 
is 3\% lower than the vector in both cases with an 
error of 1-2\%. 

Unfortunately this cannot be directly 
tested through decay modes of the $\eta_c$ 
or $\eta_b$. The decay rate 
to two photons is indirectly related to the decay 
constant as the leading term in a nonrelativistic approximation: 
\begin{equation}
\Gamma(\eta_h \rightarrow \gamma \gamma) = \frac{12 \pi e_h^4 \alpha_{QED}^2 |\psi(0)|^2}{m_h^2}.
\label{eq:etaphoton}
\end{equation}
This formula has radiative and relativistic corrections 
at the next order. 
The decay width is not known for the 
$\eta_b$ and only very poorly known 
for the $\eta_c$, with the Particle Data Group estimate 
given as 7.2(2.1) keV~\cite{pdg}. 
Substituting this into eq.~(\ref{eq:etaphoton}) 
and taking $m_c = M_{\eta_c}/2$, justifiable 
at this order, gives $f_{\eta_c}$ = 0.4(1) GeV, 
where only the large error from experiment is 
shown. This is consistent with our value but much 
less accurate so does not provide a useful test.  

As discussed earlier, a direct comparison of lattice QCD 
results for $f_{\eta_h}$ and potential model values 
for $\psi(0)$ is not particularly useful. Values for 
$\psi(0)$ for the ground state in bottomonium 
vary by a factor of 1.5 for different forms 
for the potential in~\cite{eichtenquigg}. This variation 
is reduced somewhat, and radiative corrections cancel, 
if we compare the ratio of values at $b$ and $c$. 
Here the lattice QCD result above of 1.698(14) favours the 
strong variation of $\psi(0)$ with quark mass seen in the 
Cornell potential. For this potential~\cite{eichtenquigg} gives 
a ratio $\psi_b(0)/\psi_c(0)$ of 3.1, yielding a 
decay constant ratio of 1.8. 

Figure~\ref{fig:fhh} also includes as the leftmost black point 
a value for the decay 
constant of the $\eta_s$ as determined from lattice QCD~\cite{oldr1paper}. 
Although our fit becomes unstable below 
$M$ of 2 GeV, it is interesting to see that $f_{\eta_s}$ 
does not look out of place on this plot as the light 
and heavy sectors are smoothly connected together. 

\subsection{$m_{B_s}$ and $f_{B_s}$}
\label{sec:bs}

\begin{table}
\begin{tabular}{llllll}
\hline
\hline
Set & $m_sa$ & $M_{\eta_s}a$ & $m_ha$ & $M_{H_s}a$ & $f_{H_s}a$  \\
\hline
1 & 0.061 & 0.50490(36) & 0.66 & 1.3108(6) & 0.1913(7) \\
 &  &  & 0.81 & 1.4665(8) & 0.1970(10) \\
 & 0.066 & 0.52524(36) & 0.66 & 1.3164(5) & 0.1929(7) \\
 &  &  & 0.825 & 1.4869(7) & 0.1994(10) \\
\hline
2 & 0.0492 & 0.41436(23) & 0.44 &  0.9850(4) & 0.1500(5)  \\ 
 &  &  & 0.63 &  1.2007(5) & 0.1559(7)  \\ 
 &  &  & 0.85 &  1.4289(8) & 0.1613(10)  \\ 
 & 0.0546 & 0.43654(24) & 0.44 &  0.9915(4) & 0.1516(5)  \\ 
 &  &  & 0.66 &  1.2391(5)  & 0.1586(6)  \\ 
 &  &  & 0.85 &  1.4348(7) & 0.1634(9)  \\ 
\hline
3 & 0.0337 & 0.29413(12) & 0.3 & 0.70845(17) & 0.1054(2)  \\
  & & & 0.413 &  0.84721(23) & 0.1084(2) \\
  & & & 0.7 &  1.1660(4) & 0.1112(5) \\
  & & & 0.85 &  1.3190(5) & 0.1123(6) \\
 & 0.0358 & 0.30332(12) & 0.3 & 0.71119(16) & 0.1061(2)  \\
  & & & 0.43 &  0.86982(23) & 0.1094(2) \\
  & & & 0.44 &  0.88152(23) & 0.1096(3) \\
  & & & 0.7 &  1.1684(4) & 0.1121(4) \\
  & & & 0.85 &  1.3214(5) & 0.1131(6) \\
 & 0.0366 & 0.30675(12) & 0.3 & 0.71223(16) & 0.1063(2)  \\
  & & & 0.43 & 0.87079(22) & 0.1097(2) \\
  & & & 0.44 & 0.88249(23) & 0.1099(3) \\
  & & & 0.7 &  1.1694(4) & 0.1124(4) \\
  & & & 0.85 & 1.3223(5) & 0.1135(6) \\
\hline 
4  & 0.0228 & 0.20621(19)  & 0.273 & 0.59350(24) & 0.0750(3) \\
  & & & 0.564 &  0.9313(5) & 0.0754(6) \\
  & & & 0.705 & 1.0811(8) & 0.0747(8) \\
  & & & 0.85 & 1.2279(10) & 0.0742(10) \\
\hline
5 & 0.0161 & 0.15278(28) & 0.193 &  0.43942(33) & 0.0553(4) \\
  & & & 0.5  &  0.8027(10) & 0.0541(12) \\
  & & & 0.7 &  1.0152(18) & 0.0513(22) \\
  & & & 0.85 & 1.1657(24) & 0.0495(30) \\
 & 0.0165 & 0.15484(14) & 0.195 &  0.44270(28) & 0.0555(3) \\
  & & & 0.5  &  0.8038(8) & 0.0546(11) \\
  & & & 0.7 &  1.0169(12) & 0.0526(16) \\
  & & & 0.85 & 1.1684(16) & 0.0517(21) \\
\hline
\hline
\end{tabular}
\caption{ Results for the masses and decay constants in lattice units 
of the goldstone pseudoscalars made 
from valence HISQ heavy quarks with valence HISQ 
strange quarks on the different 
MILC ensembles, enumerated in Table~\ref{tab:params}. 
Column 2 gives the $s$ mass in lattice 
units, with several values on some ensembles 
around the correctly tuned value. 
Column 3 gives the corresponding mass for 
the goldstone pseudoscalar made from the $s$ quarks, 
which is used for tuning. 
Column 4 gives the heavy quark mass. 
The corresponding values of the Naik coefficient 
are given in Table~\ref{tab:etahresults}. 
Many of these results were given earlier in~\cite{fdsupdate, bshisq}. 
}
\label{tab:bs}
\end{table}

Our calculations for heavy-strange mesons were 
described in~\cite{bshisq} and so we only 
add briefly to that discussion here. 
In Table~\ref{tab:bs} we give our full set of results,
including values 
at a variety of strange quark masses for completeness. 
In~\cite{bshisq} we used the heavy-strange mass itself as a 
proxy for the heavy quark mass and obtained good agreement 
for the mass of the $B_s$ with experiment and a value for 
$f_{B_s}$ of 225(4) MeV. 

Here, for consistency with the other calculations, 
we use instead $M_{\eta_h}$ for the heavy quark 
mass and the fit form given in eq.~(\ref{eq:fitform}). 
For the heavy-strange meson mass, as in~\cite{bshisq}, 
we fit to the mass difference: 
\begin{equation}
\Delta_{H_s,hh} = m_{H_s} - \frac{m_{\eta_h}}{2}.
\label{eq:deltahs}
\end{equation}
We take account of mistuning of the strange quark 
mass using the factor given in eq.~(\ref{eq:mstune}).  
For the decay constant fit we fix the power of 
the leading $M$-dependence, $b=-0.5$. Allowing 
$b$ to float gives results for $b$ in agreement 
with this value to within 20\%. 

\begin{figure}
\begin{center}
\includegraphics[width=0.9\hsize]{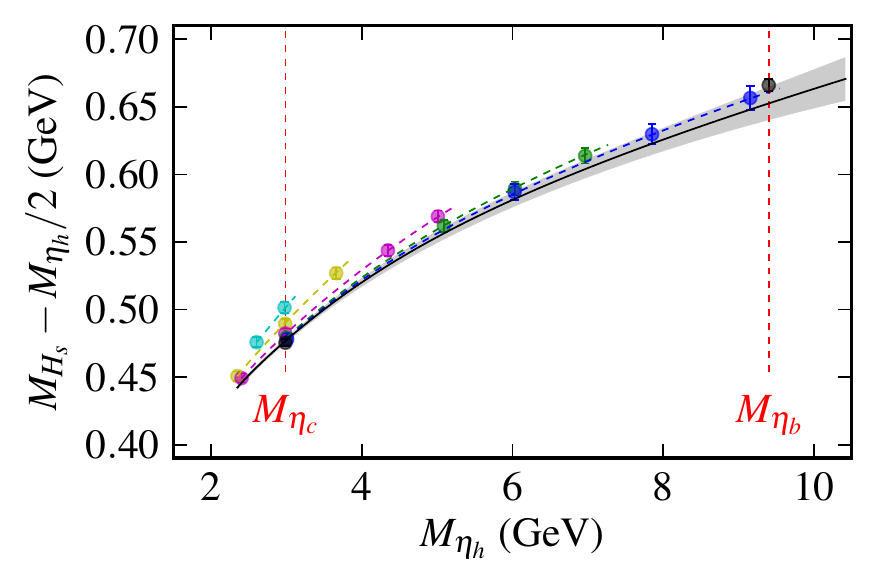}
\end{center}
\caption{Results for the difference, $\Delta_{H_s,hh}$ 
between the heavy-strange pseudoscalar meson mass and 
one half of the pseudoscalar heavyonium mass. 
Results for very coarse, 
coarse, fine, superfine and ultrafine lattices appear from left 
to right. 
The lattice QCD results have been adjusted for 
slight mistuning of the $s$ quark mass. 
The colored dashed lines give the fitted 
function for that 
lattice spacing. The black dashed line with grey error band gives 
the physical curve derived from our fit. 
The black circles with error bars at $M_{\eta_c}$ and $M_{\eta_b}$ 
are the experimental values adjusted for the 
effects from electromagnetism, $\eta_b/\eta_c$ annihilation 
and $c$ quarks in the sea, none of which is included 
in the lattice QCD calculation. } 
\label{fig:mhs}
\end{figure}

Our fit to $\Delta_{H_s,hh}$a is shown in Figure~\ref{fig:mhs} 
and gives $\chi^2=0.2$ for 
17 degrees of freedom. The values extracted at the $c$ and 
$b$ masses agree well, within $1\sigma$, with 
our earlier results~\cite{fdsupdate, bshisq}.
When account is taken of electromagnetic and 
other effects missing in the lattice calculation 
these earlier results translate into values 
for $m_{D_s} = 1.969(3) \,\mathrm{GeV}$~\cite{fdsupdate} and 
$m_{B_s} = 5.358(12) \,\mathrm{GeV}$~\cite{bshisq}. 
The increased error at the $b$ results from increased 
statistical and discretisation errors for heavier quark masses 
as well as the extrapolation in $M$. 
Our result for $m_{B_s}$ agrees within the 12 MeV 
error with that determined from full lattice QCD 
using a completely different method (NRQCD) for 
the $b$ quark~\cite{gregory} with very different 
systematic errors, providing a stringent 
test of lattice QCD.  
Our results also agree well with experiment~\cite{pdg} 
($m_{D_s} = 1.968 \,\mathrm{GeV}$ 
and $m_{B_s} = 5.367 \,\mathrm{GeV}$) 
and this provides a very strong test of QCD.

\begin{figure}
\begin{center}
\includegraphics[width=0.9\hsize]{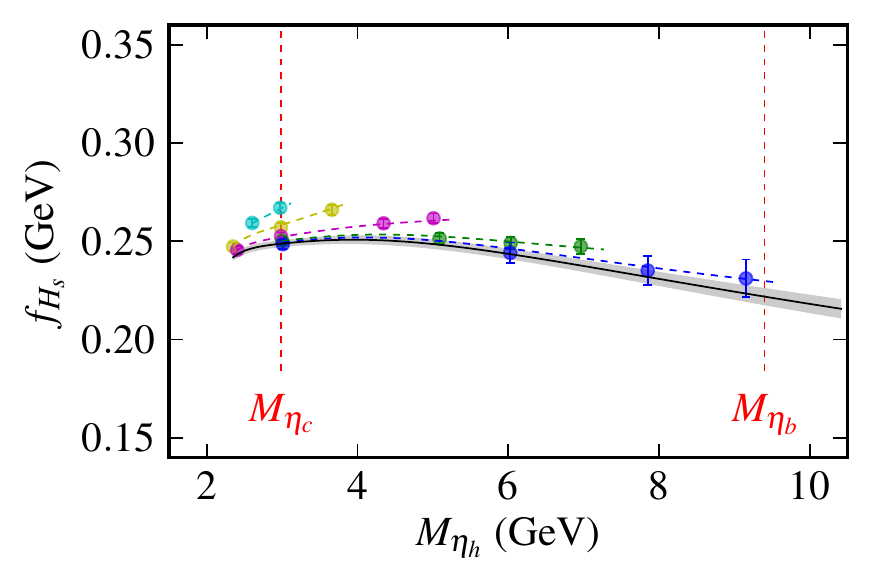}
\end{center}
\caption{Results for the pseudoscalar heavy-strange decay constant 
plotted as a function of the pseudoscalar heavyonium mass. Results 
for very coarse, 
coarse, fine, superfine and ultrafine lattices appear from left 
to right. 
The lattice QCD results have been adjusted for 
slight mistuning of the $s$ quark mass. 
The colored dashed lines give the fitted 
function for that 
lattice spacing. The black dashed line with grey error band gives 
the physical curve derived from our fit. }
\label{fig:fhs}
\end{figure}

The fit to the decay constant, $f_{H_s}$, is shown 
in Figure~\ref{fig:fhs} and gives 
$\chi^2=0.3$ for 17 degrees of freedom. 
Again results at the $b$ and $c$ agree within 
$1\sigma$ with our earlier results which 
are: $f_{D_s} = 0.2480(25) \,\mathrm{GeV}$~\cite{fdsupdate} 
and $f_{B_s} = 0.225(4) \,\mathrm{GeV}$~\cite{bshisq}. 

Figures~\ref{fig:mhs} and~\ref{fig:fhs} 
give the physical fit curves as a function of 
$M_{\eta_h}$. As expected, the curves are very similar to those 
in~\cite{bshisq} since to a large extent the change 
is simply a rescaling of the $x$-axis. 
However they provide a consistency check that the 
parameterisation we use here, taking a different 
quantity to represent the heavy quark mass, works 
just as well.

\subsection{$m_{B_c}$ and $f_{B_c}$}
\label{sec:heavycharm}

Heavy-charm mesons are of interest because a family of 
gold-plated $b\overline{c}$ mesons exists of which only 
one, the pseudoscalar $B_c$~\cite{cdfbc, d0bc}, has been seen. 
Traditionally these particles have been viewed as further 
examples, beyond $b\overline{b}$ and $c\overline{c}$, of a 
heavy-heavy system and therefore a test of our understanding 
of this area. $b\overline{c}$ mesons, however, have a lot 
in common with heavy-light systems. In fact they provide a 
bridge between heavy-heavy and heavy-light and so test 
our contol of QCD much more widely. The more accurately 
we can do these tests, the better they are. 

\begin{table}
\begin{tabular}{lllll}
\hline
\hline
Set & $m_ca$ & $m_ha$ & $M_{H_c}a$ & $f_{H_c}a$  \\
\hline
2 & 0.63 &  0.85 &  2.01651(10) & 0.2854(2)  \\ 
\hline
3  & 0.413 & 0.7 &  1.57733(7) & 0.1916(2) \\
 &  & 0.85 & 1.72373(6) & 0.2004(1)  \\
  & 0.43 & 0.7 &  1.59489(7) & 0.1938(2) \\
 & & 0.85 & 1.74105(6) & 0.2030(1)  \\
  & 0.44 & 0.7 &  1.60522(6) & 0.1952(1) \\
 &  & 0.85 & 1.75122(6) & 0.2044(1)  \\
\hline 
4  & 0.273 & 0.564 &  1.21799(8) & 0.1329(2) \\
  & & 0.705 &  1.36350(8) & 0.1367(2) \\
  & & 0.76 &  1.41872(8) & 0.1380(2) \\
  & & 0.85 & 1.50727(8) & 0.1402(2) \\
  & 0.28 & 0.564 &  1.22562(8) & 0.1338(2) \\
  & & 0.705 &  1.37103(8) & 0.1376(2) \\
  & & 0.76 &  1.42621(9) & 0.1390(2) \\
  & & 0.85 & 1.51471(9) & 0.1413(2) \\
\hline
5 & 0.195  & 0.4 &  0.90566(8) & 0.0967(3) \\
  & & 0.5  &  1.01457(9) & 0.0985(4) \\
  & & 0.7 &  1.22392(10) & 0.1005(4) \\
  & & 0.85 & 1.37366(10) & 0.1018(5) \\
\hline
\hline
\end{tabular}
\caption{ Results for the masses and decay constants in lattice units 
of the goldstone pseudoscalars made 
from valence HISQ heavy quarks with valence HISQ 
charm quarks on the different 
MILC ensembles, enumerated in Table~\ref{tab:params}. 
Set 1 is missing because $m_ca$ is already 
close to the highest heavy quark mass that we use. 
Column 2 gives the $c$ mass in lattice 
units, with several values on some ensembles 
around the tuned $c$ mass, and column 3 the heavy quark mass. 
The corresponding values of the Naik coefficient 
are given in Table~\ref{tab:etahresults}. 
}
\label{tab:bc}
\end{table}

Lattice QCD calculations of the $B_c$ mass can be done 
very accurately. Indeed the mass of 
the $B_c$ was predicted ahead of experiment with a 
22 MeV error~\cite{allison} using NRQCD for the 
$b$ quark and the `Fermilab' clover action for the 
$c$ quark. 
The error was later 
reduced to 10 MeV by using a more highly improved 
action, HISQ, for the charm quark~\cite{gregory}.  
Here we use the HISQ action for both the $c$ quark  
and the heavier quark up to the $b$ mass
to obtain results in a completely different heavy quark formalism. 
In addition we calculate the decay constant of the $B_c$ for 
the first time in full lattice QCD. 

\begin{figure}
\begin{center}
\includegraphics[width=0.9\hsize]{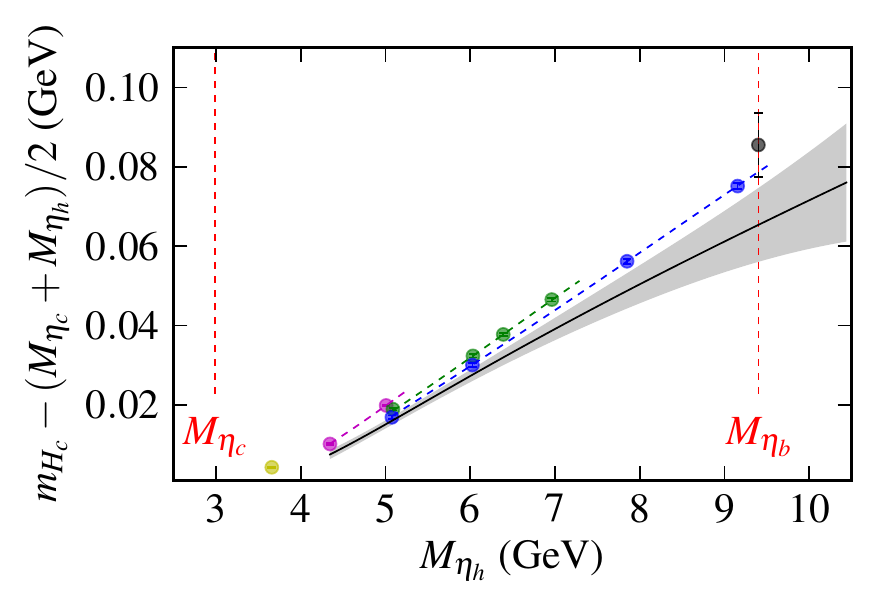}
\end{center}
\caption{Results for the mass difference between 
the $H_c$ meson and the average of the associated 
heavyonium pseudoscalar meson masses 
plotted as a function of the pseudoscalar heavyonium mass. Results 
for coarse, fine, superfine and ultrafine lattices appear from left 
to right. 
The lattice QCD results have been adjusted for 
slight mistuning of the $c$ quark mass. 
The colored dashed lines give the fitted 
function for that 
lattice spacing. The black line with grey error band gives 
the physical curve derived from our fit. 
The black circle with error bar at $M_{\eta_b}$ 
gives the experimental value adjusted for the 
effects from electromagnetism, $\eta_b/\eta_c$ annihilation 
and $c$ quarks in the sea, none of which is included 
in the lattice QCD calculation. } 
\label{fig:deltabc}
\end{figure}

\begin{figure}
\begin{center}
\includegraphics[width=0.9\hsize]{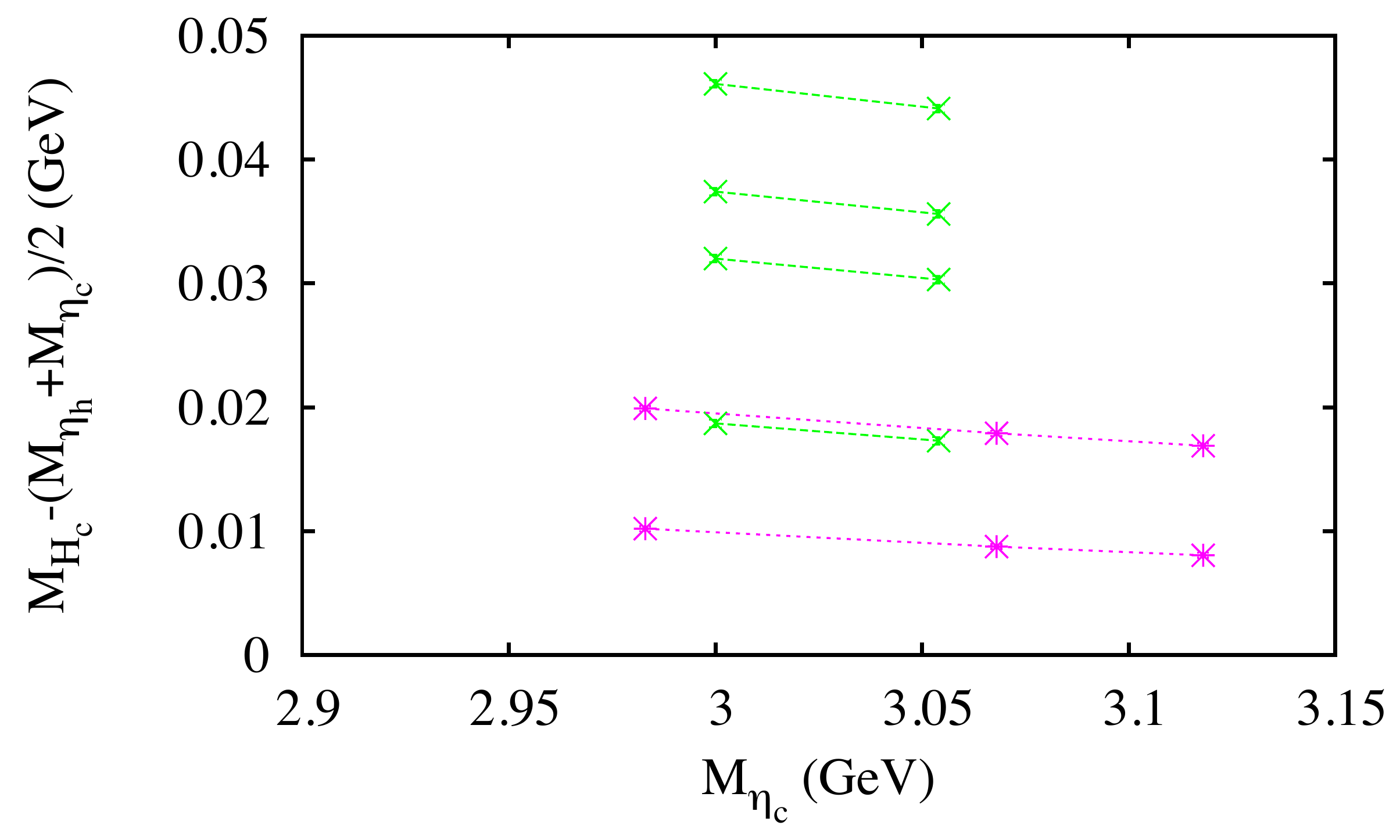}
\end{center}
\caption{Results for the mass difference between 
the $H_c$ meson and the average of the associated 
heavyonium pseudoscalar meson masses 
plotted as a function of the pseudoscalar charmonium mass. Results 
are given for two heavy quark masses on 
fine lattice set 3 (pink bursts) and 
four heavy quark masses on superfine lattices set 4 (green crosses).
Lines are drawn to guide the eye. }
\label{fig:mhcmc}
\end{figure}

To determine the $B_c$ mass we use the 
mass difference to the average of 
the associated heavyonium states: 
\begin{equation}
\Delta_{H_c,hh} = M_{H_c} - \frac{1}{2}(M_{\eta_c}+M_{\eta_h}).
\end{equation}
$\Delta_{H_c,hh}$ is a measure of the difference in 
binding energy between the symmetric heavyonium states made 
of $c$ and $h$ quarks and the heavyonium state made 
of two different mass quarks, $c$ and $h$.   
Here we map out $\Delta_{H_c,hh}$ as a function of the 
heavy quark mass, and reconstruct $M_{B_c}$ from 
$\Delta_{H_c,hh}$ determined at $h=b$. 
$\Delta_{H_c,hh}$ can be determined with high statistical accuracy 
because all of the states involved have very little noise. 
The fact that $\Delta_{H_c,hh}$ is very small 
( 0 for $m_h = m_c$ by definition and less than 100 MeV 
when $m_h = m_b$) 
also means that lattice errors from, for example, the uncertainty 
in the lattice spacing are very small. 
In fact, for this calculation, as discussed below, 
key sources of error are
the uncertainties from electromagnetic, annihilation 
and $c$-in-the-sea shifts to the masses. 

Table~\ref{tab:bc} gives our results for the masses 
and decay constants of the $H_c$ mesons calculated 
using quark masses that are close to that of 
the $c$ quark mass on 
each ensemble and then all the heavier masses for $h$.  
We give results for more than one value 
of the $c$ quark mass on the fine and superfine 
ensembles (sets 3 and 4) so that slight mis-tuning 
in the $c$ quark mass can be corrected for. 
It is clear from the results that $\Delta_{H_c,hh}$ can be calculated 
with a statistical accuracy of better than 1 MeV. 
Errors from uncertainties in the lattice spacing are 
also at this level. 

Figure~\ref{fig:deltabc} shows $\Delta_{H_c,hh}$ plotted against 
$M_{\eta_h}$ for the results at different values of 
the lattice spacing. 
A fairly clear linear dependence is evident. 
$\Delta_{H_c,hh}$ would be expected to
increase linearly with $M_{\eta_h}$ at large $M_{\eta_h}$, 
in the same way as $\Delta_{H_s,hh}$,
from a simple potential model argument. The 
binding energy of the $\eta_h$  becomes increasingly 
negative, roughly in proportion to $M_{\eta_h}$ as it 
increases (at least for a $r^N$ potential with $N=-1$), 
whilst the binding energy 
of the $H_c$ meson does not change. 
A corollary of this is that 
the dependence of $\Delta_{H_c,hh}$ on $M_{\eta_c}$ 
(as proxy for $m_c$) would also 
then be expected to be linear with a slope of 
opposite sign and roughly three times the magnitude. 
The factor of three is because the binding energy
of the $\eta_c$ becomes more negative as $M_{\eta_c}$ 
increases, with the same dependence as the 
$\eta_h$ binding energy has on $M_{\eta_h}$. 
The $H_c$ binding energy will also become 
more negative but have double the slope 
because the reduced mass of the $H_c$ system is 
roughly $m_c$ rather than $m_c/2$ for the $\eta_c$. 
On top of this $M_{\eta_c}$ appears halved in 
$\Delta_{H_c,hh}$. 

Interestingly this factor of -3 does seem 
to be approximately true in comparing Figure~\ref{fig:mhcmc}, 
which shows the dependence of $\Delta_{H_c,hh}$ on $M_{\eta_c}$,
with Figure~\ref{fig:deltabc}. 
Figure~\ref{fig:deltabc} gives a slope of $\approx 0.012$ 
(over the full range) and Figure~\ref{fig:mhcmc} gives slopes
varying from -0.03 to -0.04 over a small range in $M_{\eta_c}$, 
as $M_{\eta_h}$ increases. In our fit to $\Delta_{H_c,hh}$ we include 
the effect of mistuning $m_c$ (from eq.~(\ref{eq:mctune}))    
and obtain consistent values from that. 

We fit $\Delta_{H_c,hh}$ as a function of 
$M_{\eta_h}$ (above 4 GeV) using the fit form described 
in section~\ref{sec:latt}. The leading mass 
dependence is taken to be $M_{\eta_h}-M_{\eta_c}$, 
so that $\Delta_{H_c,hh}$ vanishes when 
$M_{\eta_h}=M_{\eta_c}$ as it must by definition. 
As described in section~\ref{sec:latt} 
we include a sum of power correction terms and 
lattice spacing dependent terms with priors 
given in Table~\ref{tab:f0}. 
The fit gives $\chi^2$ of 0.3 for 11 degrees of 
freedom and result:
\begin{equation}
\Delta_{B_c,bb} = 0.065(9) \mathrm{GeV}.
\label{eq:bcres}
\end{equation}

The resulting physical curve of 
heavy quark mass dependence 
is shown in grey on Figure~\ref{fig:deltabc}.
The comparison to experiment is given by the 
black dot with error bar at $h=b$. 
This experimental result has been shifted 
to be the appropriate value to compare to 
our lattice QCD calculation as we now describe. 
The current world-average experimental result for 
$M_{B_c} - 0.5(M_{\eta_c}+M_{\eta_b})$ 
is 92(6) MeV~\cite{pdg}. There is a 
sizeable experimental error coming 
mainly from the $B_c$ but also from the $\eta_b$. 
Our lattice QCD calculation is done 
in a world without electromagnetism or 
$c$ quarks in the sea and in which the 
$\eta_b$ and $\eta_c$ do not annihilate. 
The absence of these effects (i.e. to compare to our lattice 
result) produces shifts to the masses as 
discussed in section~\ref{sec:latt}. Estimated values 
for the shifts 
are given in Table~\ref{tab:adjust}. 
The net effect is to shift the experimental 
value of $\Delta_{B_c,hh}$ down by -8(7) MeV, where the 
error takes the shifts to be correlated. 
The `experimental' value of $\Delta_{B_c,hh}$ to compare to our lattice 
result is then 84(9) MeV, marked on Figure~\ref{fig:deltabc}. 
Our lattice result agrees with experiment, once these 
shifts are made, within $2\sigma$. 

From $\Delta_{B_c,bb}$ we can reconstruct the 
$B_c$ meson mass, now applying the shifts above 
to the lattice QCD calculation to obtain a result
that can be compared to experiment. 
This gives the result
\begin{equation}
M_{B_c} = 6.259(9)(7) \mathrm{GeV}.
\label{eq:mbc}
\end{equation}
Here the first error comes from the fit and the 
second error from the shifts applied to include 
missing real world effects as well as experimental 
uncertainties in the $\eta_b$ and $\eta_c$ masses. 
As can be seen, this 
is a sizeable part of the total error in this case. 
We also include in this second error an estimate 
of sea quark mass effects using results from~\cite{fdsupdate}. 
There we saw no such for an equivalent quantity 
for $m_{D_s}$ within 1 MeV statistical errors and 
so take that as the error here. 
Table~\ref{tab:bcerrors} gives the complete error budget 
for $\Delta_{B_c,bb}$ breaking down both errors 
into their components. 

\begin{table}
\begin{tabular}{lccc}
\hline
\hline
Error & $\Delta_{B_c,bb}$ & $f_{B_c}$ & $\Delta_{B_c, bs}$ \\
\hline
statistics  & 8.4 &  0.7 & 0.5 \\
$M$ extrapoln  & 3.1  & 0.2 & 0.2 \\
$a^2$ extrapoln  & 10.9 & 0.7 & 0.4 \\
$r_1$  & 0.7 & 0.6 & 0.3 \\
$r_1/a$  & 1.4 & 0.8 & 0.3 \\
$M_{\eta_{c}}$ & 0.9 & 0.5 & 0.3 \\
sea quark mass effects & 1.5 &  0.1 & 0.1 \\
electromagnetism & $3.1^*$ & 0.4 & $0.2^*$ \\
$c$ in the sea & $5.3^*$ & 0.04 & $0.1^*$ \\
$\eta_{b,c}$ annihiln & $2.7^*$ & - & - \\
\hline
Total (\%) & 18 & 1.6 & 0.9 \\
\hline
\hline
\end{tabular}
\caption{ Full error budget for $\Delta_{B_c, bb}$, 
$f_{B_c}$ and $\Delta_{B_c, bs}$ given as a percentage of the value. 
See the text for a fuller 
description of each error. The total error is 
obtained by adding the individual errors in quadrature, 
except for the final three systematic errors (starred) for 
$\Delta_{B_c,bb}$ and $\Delta_{B_c, bs}$ which 
are correlated and so simply 
added together before being combined in quadrature with 
the others.  
 }
\label{tab:bcerrors}
\end{table}

Our result for $M_{B_c}$ can be compared to 
experiment (6.277(6) GeV) and to our result 
from lattice QCD using a completely different 
formalism, NRQCD, for the $b$ quark 
(6.280(10) GeV~\cite{gregory}). We agree, within 
$2\sigma$ with both results even allowing for the 
fact that the comparison within lattice QCD can be 
done before any shifts are made or errors allowed for them.  
This is a strong confirmation of the control over 
errors that we now have in lattice QCD. 

The method given here for determining $m_{B_c}$ 
(as for the method for $m_{B_s}$ in section~\ref{sec:bs})
does depend on the experimental $\eta_b$ mass; 
the mass difference determined in lattice QCD is 
not particularly sensitive to it but when the 
mass is reconstructed from the difference, $m_{\eta_b}/2$ 
is added in. Recent results from the Belle 
collaboration~\cite{belleetab} have 
$M_{\eta_b}=9.402(2) \,\mathrm{GeV}$, significantly 
higher than the previous world-average~\cite{pdg}. 
Using the Belle result for $M_{\eta_b}$ pushes our 
values for $m_{B_c}$ and $m_{B_s}$ 6 MeV higher.
In both cases this improves the agreement
with experiment but is not significant given the 
11 MeV error. Note that our earlier NRQCD results 
are hardly affected at all by a change in 
the $\eta_b$ mass because they determined a mass 
difference to the spin-average of the $\Upsilon$ 
and $\eta_b$ masses, which is dominated by the 
$\Upsilon$ mass. 

\begin{figure}
\begin{center}
\includegraphics[width=0.9\hsize]{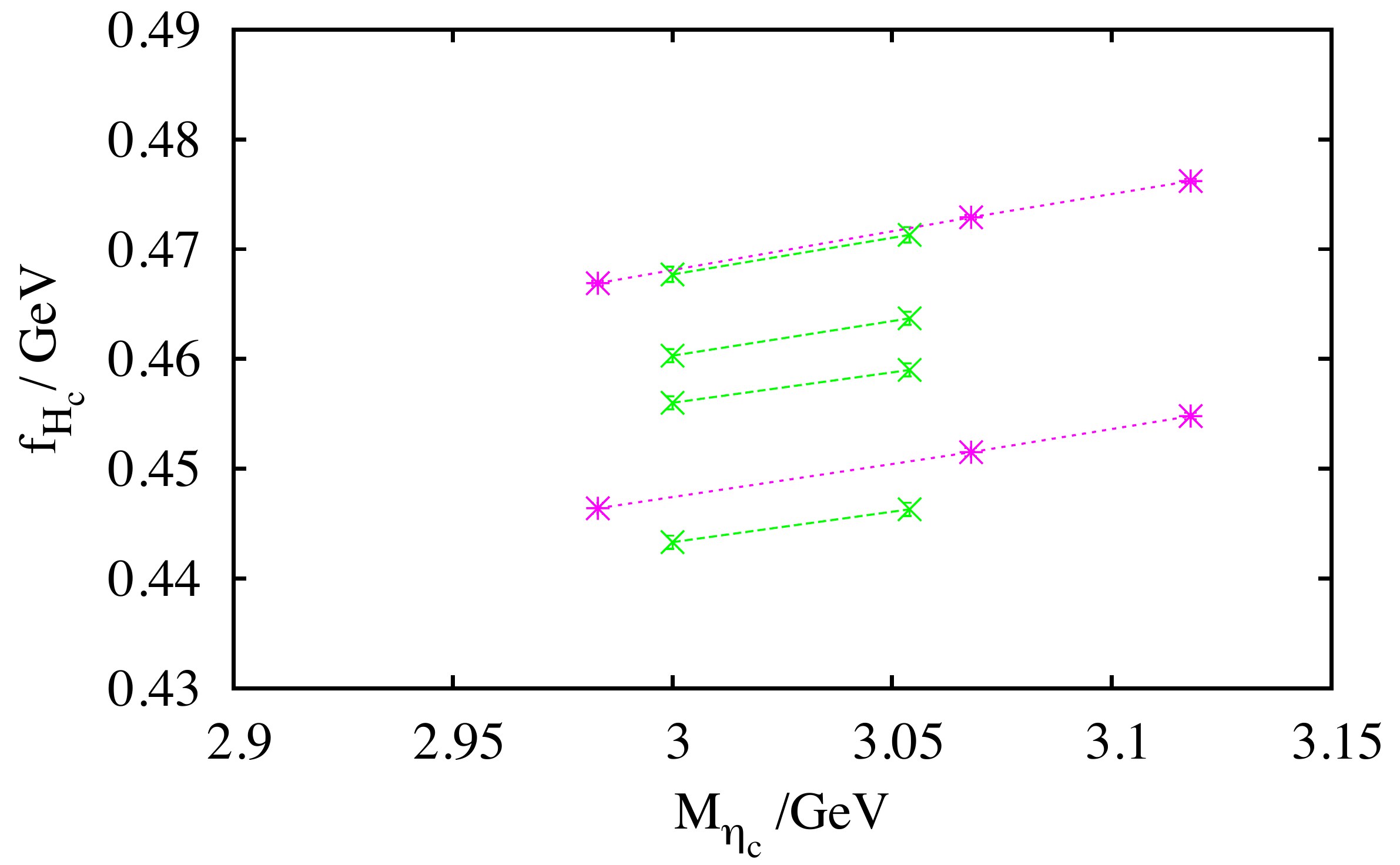}
\end{center}
\caption{Results for the heavy-charm decay constant 
plotted as a function of the $c$ quark mass, given by
the mass of the $\eta_c$ meson. 
Results are given for multiple heavy quark 
masses on fine lattices (pink bursts) and superfine 
lattices (green crosses). Lines are drawn to guide the eye. 
}
\label{fig:fhcmc}
\end{figure}

\begin{figure}
\begin{center}
\includegraphics[width=0.9\hsize]{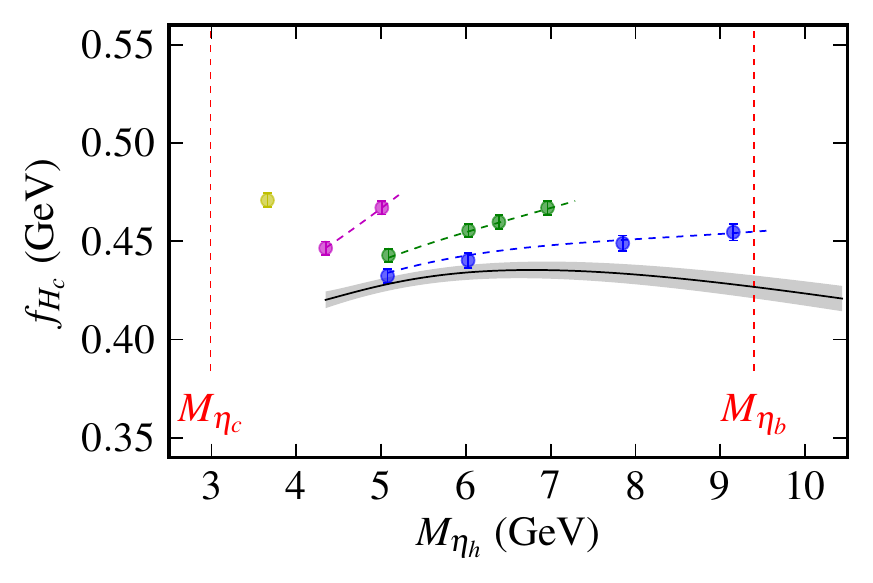}
\end{center}
\caption{Results for the heavy-charm decay constant 
plotted as a function of the pseudoscalar heavyonium mass. Results 
for coarse, fine, superfine and ultrafine lattices appear from left 
to right. 
The lattice QCD results have been adjusted for 
slight mistuning of the $c$ quark mass. 
The colored dashed lines give the fitted 
function for that 
lattice spacing. The black line with grey error band gives 
the physical curve derived from our fit. 
}
\label{fig:fhc}
\end{figure}

Results for the $H_c$ decay constant, $f_{H_c}$, are also 
given in Table~\ref{tab:bc}. The rate for $B_c$ 
leptonic decay to $l\nu$ via a $W$ boson is 
proportional to the square of the decay constant 
multiplied by CKM element $V_{cb}$ as in eq.~(\ref{eq:gamma}). 
In practice this 
decay will be very hard to see experimentally, 
but a lattice QCD calculation of the decay constant 
also provides a useful test for phenomenological 
model calculations. 

The results at 
different values of $m_c$ can again be used to 
tune the decay constant accurately to the 
result at the physical $c$ quark mass. 
Figure~\ref{fig:fhcmc} shows the dependence 
of $f_{H_c}$ on $M_{\eta_c}$ acting as a proxy for 
the $c$ quark mass. Results on fine and superfine 
lattices are shown -- there is clear agreement on 
the physical slope of $f_{H_c}$ with $M_{\eta_c}$ 
between superfine and fine and it does not vary 
with the heavy quark mass. 
The slope is small, approximately 0.06, but clearly 
visible. We will compare this to the slope 
for $f_{H_s}$ with $m_s$ in section~\ref{sec:discussion}. 

The $H_c$ decay constant is plotted as a function 
of $M_{\eta_h}$ in Figure~\ref{fig:fhc}. Notice that 
it is much flatter than the corresponding plot 
for $f_{\eta_h}$ (Figure~\ref{fig:fhh}). 
We expect behaviour as $1/\sqrt{M_{\eta_h}}$ whether 
we view heavy-charm as a heavy-light system (in which 
case the behaviour will be similar to heavy-strange) 
or as a heavy-heavy system (in which case the 
argument becomes that $\psi(0)$
depends on the reduced mass $\mu$, tending to $m_c$ for 
large $m_h$, and then the decay constant falls 
as the square root of the heavy mass). 

As before, we fit $f_{H_c}$ to the function of $M_{\eta_h}$ 
(above 4 GeV) described in section~\ref{sec:latt}. 
We take the leading term given in Table~\ref{tab:f0} to be 
that expected from HQET arguments appropriate to heavy-light 
physics. 
Our fit has $\chi^2 = 0.7$ for 11 degrees of freedom 
and gives result: 
\begin{equation}
f_{B_c} = 0.427(6)(2) \mathrm{GeV}. 
\end{equation} 
Here the first error is from the fit and the second from 
additional systematic effects missing from our lattice 
QCD calculation. These we estimate based on the arguments
given for the $\eta_h$ in section~\ref{sec:etab}. The 
error from missing electromagnetism and from sea quark 
mass effects we take to be the same as for the $\eta_b$  
at 0.4\% and 0.1\% respectively; 
missing $c$ in the sea should be a factor of $m_c/m_b$ 
smaller at 0.04\%. Table~\ref{tab:bcerrors} gives the 
complete error budget. 

$f_{B_c}$ can be converted into a branching fraction for leptonic 
decay using the formula of eq.~(\ref{eq:gamma}) and 
the unitarity value of $V_{cb}$. We predict a branching 
fraction to $\tau \nu$ of $0.0194(18)$. 
The error here comes mainly from the 
experimental determination of the $B_c$ lifetime with 
a smaller effect from the uncertainty in $V_{cb}$. 
Our value for $f_{B_c}$ contributes a 3\% error. 
Because of helicity suppression the branching fraction 
smaller for other lepton final states 
($8 \times 10^{-5}$ to $\mu \nu$, 
for example).  

The value we obtain for $f_{B_c}$ can be compared to 
results from potential models. As discussed earlier in the 
context of $f_{\eta_h}$, potential model results have a lot of 
variability and raw values for $\psi(0)$ need renormalisation. 
A more useful comparison is to compare ratios. 
Our lattice QCD results give $f_{B_c}/f_{\eta_c} = 1.08(1)$ 
and $f_{\eta_h}/f_{B_c} = 1.57(2)$. 
The range of potentials considered in~\cite{eichtenquigg} 
give values from 0.90 to 1.02 for $f_{B_c}/f_{\eta_c}$ and 
1.34 to 1.72 for $f_{\eta_h}/f_{B_c}$. Again 
the largest number is always from the Cornell potential. 
Potential model values for $\psi(0)$ converted to 
$f_{B_c}$ simply using $f=\psi(0)\sqrt{12/M_{B_c}}$
yield results varying from 0.5 to 0.7 GeV i.e. significantly 
larger than the well-defined value for $f_{B_c}$ from lattice QCD. 

The values for $f_{B_c}$ from potential models provide input 
to estimates of the production cross-section of the 
$B_c$ at the LHC. In the factorisation approach the cross-section 
is proportional to the square of $f_{B_c}$, with typical 
values for $f_{B_c}$ being taken as 0.48 GeV~\cite{chang}.  
Our results indicate that this could be leading to 
a 25\% overestimate of the production rate. 

\section{Discussion}
\label{sec:discussion}

An interesting issue is to what extent the $B_c$ meson is a 
heavy-heavy particle and to what extent, a heavy-light one at 
the physical values we have for $b$ and $c$ quark masses. 
Here we address this by comparing the behaviour of $B_c$ 
properties to those of $\eta_h$ and $B_s$ using the 
results from section~\ref{sec:results}. 

An alternative to calculating $\Delta_{H_c,hh}$ to study 
the heavy-charm meson mass is to take differences 
between heavy-charm and heavy-strange and charm-strange 
mesons. We define
\begin{equation}
\Delta_{H_c,hs} = M_{H_s} + M_{D_s} - M_{H_c},
\end{equation}
so that $\Delta_{H_c,hs}$ is a positive quantity. 
Once again it amounts to a difference in binding 
energies but now between a set of mesons that 
are all effectively `heavy-light' states. Indeed 
a study of $\Delta_{H_c,hs}$ shows us to what 
extent the $B_c$ can be considered a heavy-light 
particle rather than, or as well as, a heavy-heavy one. 

\begin{figure}
\begin{center}
\includegraphics[width=0.9\hsize]{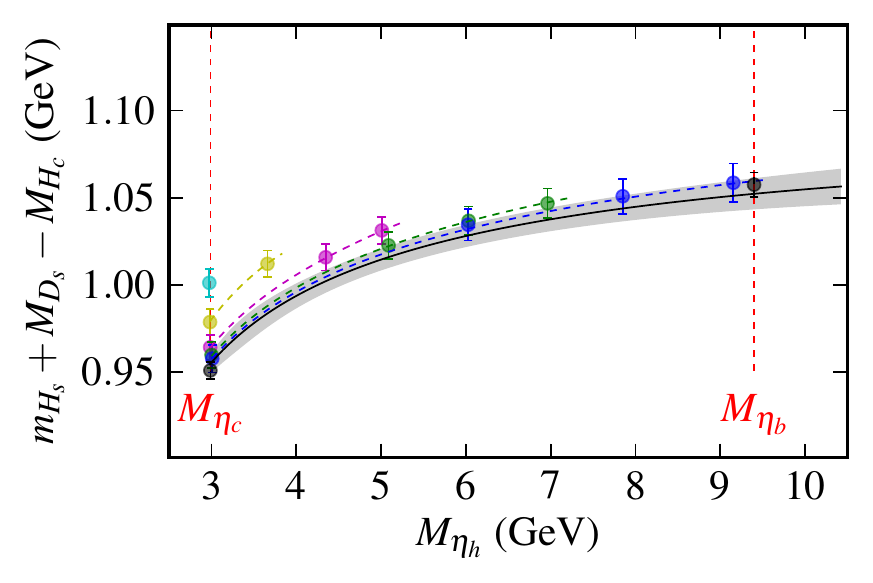}
\end{center}
\caption{Results for the mass difference between the heavy-charm 
meson and the corresponding heavy-strange and charm-strange 
mesons 
plotted as a function of the pseudoscalar heavyonium mass. Results 
for coarse, fine, superfine and ultrafine lattices appear from left 
to right. 
The lattice QCD results have been adjusted for 
slight mistuning of the $c$ and $s$ quark masses. 
The colored dashed lines give the fitted 
function for that 
lattice spacing. The black line with grey error band gives 
the physical curve derived from our fit. 
The black circles with error bars at $M_{\eta_b}$ 
and $M_{\eta_c}$
give experimental values adjusted for the 
effects from electromagnetism, $\eta_b/\eta_c$ annihilation 
and $c$ quarks in the sea, none of which is included 
in the lattice QCD calculation. } 
\label{fig:mhcs}
\end{figure}

Figure~\ref{fig:mhcs} shows $\Delta_{H_c,hs}$, with all 
results tuned accurately to the correct $c$ and $s$ 
masses, as a function of the heavy quark mass, again 
given by the $\eta_h$ mass. 
In fact $\Delta_{H_c,hs}$ shows very little dependence 
on the heavy quark mass above a value of $M_{\eta_h}$ 
of about 6 GeV.  HQET would 
expect the leading $m_h$-dependent piece of 
$\Delta_{B_c,hs}$ to be given by the difference of 
the expectation values of the kinetic energy operator,
$p_h^2/2m_h$, 
for the heavy quark in a heavy-charm meson and 
a heavy-strange meson, ignoring the effect of 
spin-dependent terms which are expected to be smaller. 
Figure~\ref{fig:mhcs} shows that this difference 
is not large i.e. the charm quark 
is behaving in a similar way to a light quark (but does have a larger 
expectation value for its kinetic energy operator 
as might be expected) 
when combined with a 
heavy quark of order twice its mass or heavier. 

We fit $m_{H_c,hs}$ as described in section~\ref{sec:latt} 
and using the fit form and priors tabulated in Table~\ref{tab:f0}. 
Our fit has $\chi^2$ of 0.3 
for 14 degrees of freedom. It returns the coefficient of 
the first term in $M_{\eta_h}^{-1}$ 
as $-0.4(8) \mathrm{GeV}/M_{\eta_h}$.
This quantifies the statement made above about the 
slope of $1/m_h$ corrections. The 
coefficient is not very accurately determined 
because we allow for many higher order terms. 
In fact the sign of the slope is clear from 
Figure~\ref{fig:mhcs} with a positive slope 
with $M_{\eta_h}$ corresponding to a negative 
value for the coefficient of the $1/M$ term, as 
expected. 

The variation of $\Delta_{H_c,hs}$ with 
$M_{\eta_c}$ agrees well with that found 
in our calculation using NRQCD $b$ quarks~\cite{gregory}
giving a slope of 0.07 at the $b$. 
Likewise the variation with $M_{\eta_s}^2$ also 
agrees well with the slope of 0.4 found in~\cite{gregory}. 

Our fit to $\Delta{H_c,hs}$ is independent of 
our earlier fit to $\Delta_{H_c,hh}$ 
(although it uses some of the same numbers)
and so the results provide a consistency check. We 
find at $h=b$ that:
\begin{equation}
\Delta_{B_c,bs} = 1.052(9)(3) \mathrm{GeV}
\end{equation}
which agrees well within $1 \sigma$ with the same 
quantity calculated using NRQCD $b$ quarks~\cite{gregory}. 
The result when $h=c$ is consistent within $1\sigma$ 
with double the result from $m_{D_s}-m_{\eta_c}/2$ given 
in~\cite{fdsupdate}. 
The first error above is from the fit and the second from 
the systematic error for sea quark mass effects, taking the 
same 1 MeV as for $\Delta_{B_c, bb}$, and 
the effects of missing electromagnetism 
and $c$ in the sea. The shifts and errors for these latter  
effects are given in Table~\ref{tab:adjust} and we take 
those errors to be correlated. 
The value above for $\Delta_{B_c,bs}$ combined with experimental 
results for $M_{B_s}$ and $M_{D_s}$~\cite{pdg} (the net shift 
from Table~\ref{tab:adjust} amounts to a negligible 0.2 MeV) 
gives:
\begin{equation}
M_{B_c} = 6.285(9)(3) \mathrm{GeV}, 
\label{eq:mbchs}
\end{equation}
consistent within $2 \sigma$ with our result 
from $\Delta_{B_c,bb}$ given 
in section~\ref{sec:results}, and slightly more accurate. 
We therefore adopt it as our final result here. 
The complete error budget for   
$\Delta_{B_s, bs}$ is given in Table~\ref{tab:bcerrors}. 

\begin{figure}
\begin{center}
\includegraphics[width=0.9\hsize]{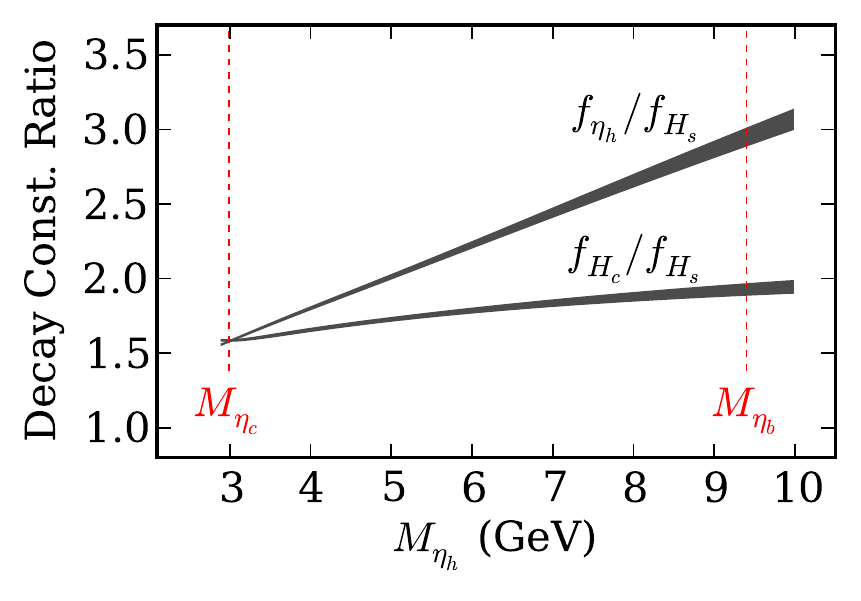}
\end{center}
\caption{Results for the ratio of pseudoscalar decay constants, 
heavy-charm and heavy-heavy to heavy-strange
plotted as a function of the pseudoscalar heavyonium mass. 
The results are obtained from the physical curves given 
in Figures~\ref{fig:fhh},~\ref{fig:fhs} and~\ref{fig:fhc}. 
}
\label{fig:fhcs}
\end{figure}

In figure~\ref{fig:fhcs} we show the ratio of $\eta_h$ and 
$H_c$ decay constants to that of the $H_s$, plotted from 
our physical curves as a function of $M_{\eta_h}$. The ratio 
$f_{\eta_h}/f_{H_s}$ rises strongly with $M_{\eta_h}$,
because of the big difference in the dynamics of heavy-heavy 
and heavy-strange mesons,
whereas the ratio $f_{H_c}/f_{H_s}$ tends to a constant 
at large $M_{\eta_h}$. As explained in section~\ref{sec:heavycharm} 
this latter behaviour would be expected whether the heavy-charm is viewed 
as a heavy-heavy or heavy-light state, because the reduced 
mass of the heavy-charm system is controlled by the charm 
mass in the large heavy mass limit. 

\begin{figure}
\begin{center}
\includegraphics[width=0.9\hsize]{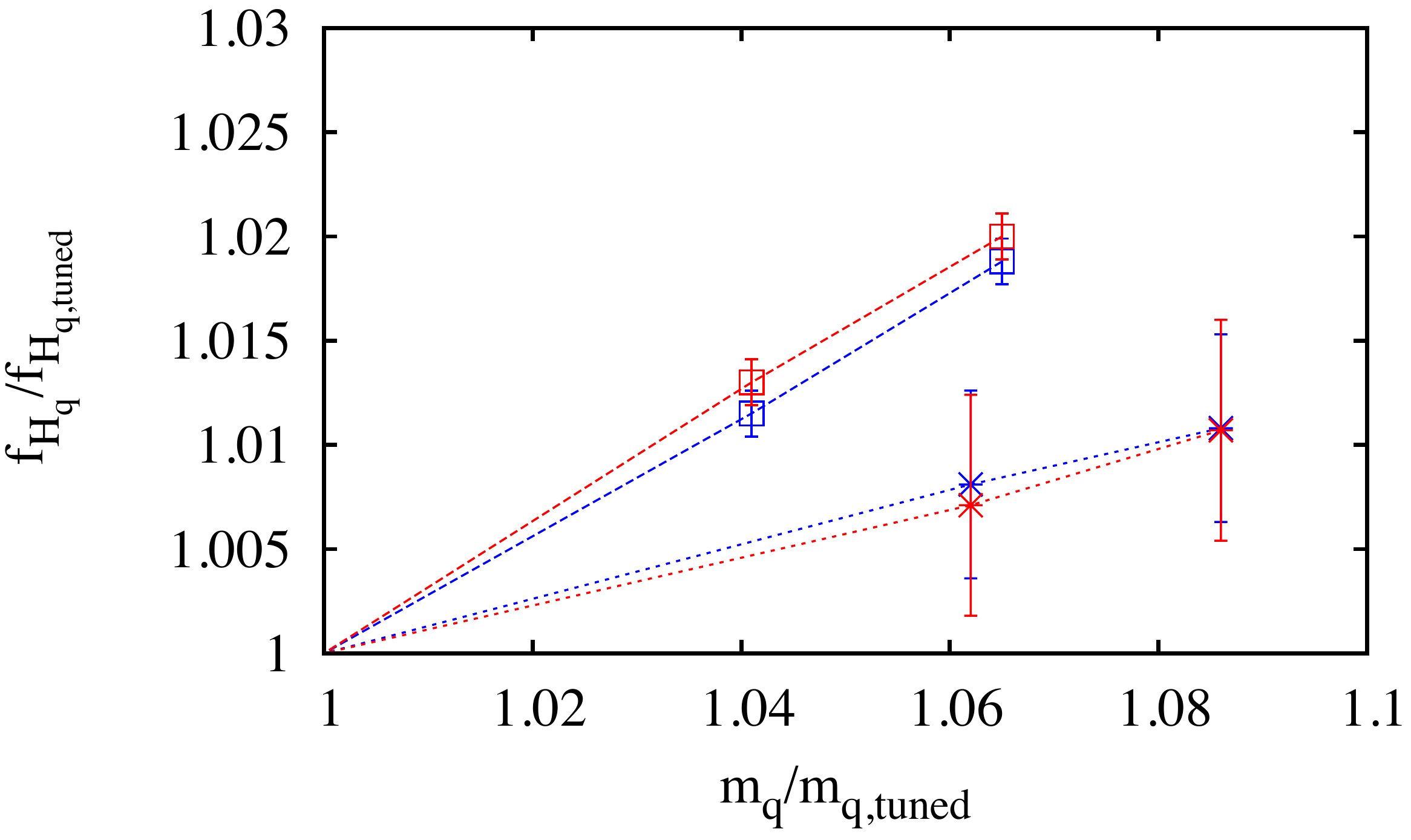}
\end{center}
\caption{Comparison of the effect of `detuning' the charm and 
strange quark masses on the heavy-charm and heavy-strange 
decay constants. Open squares show the fractional change in 
$f_{H_c}$ for a given fractional change in $M_{\eta_c}$ (as 
proxy for $m_c$) for two different heavy quark masses (in 
blue $m_ha=0.7$ and red $m_ha=0.85$) on the fine lattices set 3. 
Burst show the fractional change in $f_{H_s}$ for a 
given fractional change in $M_{\eta_s}^2$ (as proxy for 
$m_s$) for the same two heavy quark masses on set 3.   
Lines are drawn to guide the eye. 
}
\label{fig:fhqcomp}
\end{figure}

Further insight comes from comparing the dependence of 
the heavy-charm and heavy-strange 
decay constants on $m_c$ and $m_s$ respectively.
Figure~\ref{fig:fhqcomp} plots the relative change of $f_{H_c}$ 
or $f_{H_s}$ to its value at the tuned mass point for a given 
relative change in the light quark mass. The strange quark mass is 
monitored by the value of $M_{\eta_s}^2$, the charm mass 
by $M_{\eta_c}$. The results come from the fine lattices, set 3,
where we have multiple $m_c$ and $m_s$ values close 
to the tuned point. Results are 
plotted for two values of the heavy quark mass, $m_ha=0.7$ 
and $m_ha=0.85$ but little difference between them is seen. 

The dependence of $f_{H_s}$ on $m_s$ is not very 
strong~\cite{bshisq}, as expected since $f_{H_s}$ 
and $f_{H}$ differ only by around 20\% for a change 
by a factor of 27 in light quark mass. The dependence 
of $f_{H_c}$ on $m_c$ is larger by about a factor 
of two. However the slope of the Figure~\ref{fig:fhqcomp} 
is 1/3 (see also Figure~\ref{fig:fhcmc}), much less than the 
slope of 1 expected if $f_{B_c} \propto m_c$. This latter 
behaviour would be approximately that expected in a heavy-heavy 
picture in which $\psi(0) \propto \mu$, with the reduced 
mass, $\mu$, close to $m_c$ in the $B_c$ case. The linear 
behaviour of $\psi(0)$ would be consistent with the 
picture we have of the $\eta_h$ in Figure~\ref{fig:fhh}, 
where $\mu \approx M_{\eta_h}/4$, using $b\approx 0.5$. 

\begin{table}
\begin{tabular}{ccccccc} \hline \hline
$M_{\eta_h}$ & $f_{\eta_h}$ & $f_{H_s}$ & $f_{H_c}$ & $\Delta_{H_s,hh}$ & $\Delta_{H_c,hh}$ & $\Delta_{H_c,hs}$ \\ \hline
3 &            0.394(2)     & 0.249(2)  & --        & 0.477(2)          & 0.000(0)          & 0.956(6) \\
4 &            0.452(2)     & 0.251(2)  & 0.417(6)  & 0.520(3)          & 0.004(1)          & 0.994(7) \\
5 &            0.501(3)     & 0.249(2)  & 0.427(3)  & 0.554(4)          & 0.015(1)          & 1.014(6) \\
6 &            0.546(4)     & 0.244(3)  & 0.434(4)  & 0.581(6)          & 0.027(2)          & 1.028(6) \\
7 &            0.586(4)     & 0.237(3)  & 0.435(4)  & 0.605(7)          & 0.039(3)          & 1.038(7) \\
8 &            0.623(5)     & 0.231(4)  & 0.433(5)  & 0.626(9)          & 0.050(5)          & 1.045(8) \\
9 &            0.655(6)     & 0.224(4)  & 0.429(6)  & 0.645(11)         & 0.061(8)          & 1.050(8) \\
\hline
\end{tabular}
\caption{Values for the various quantities that we fit here 
evaluated at masses, $M_{\eta_h}$, between that of $c$ and $b$.
These are obtained from our fit functions at $a=0$ and tuned 
$s$ and $c$ masses. 
All numbers are in~GeV. There is no result for $f_{H_c}$ at 
3 GeV because that point is not included in that fit. 
}
\label{tab:fitnums}
\end{table}

\section{Conclusions}
\label{sec:conclusions}

\begin{figure}
\begin{center}
\includegraphics[width=0.9\hsize]{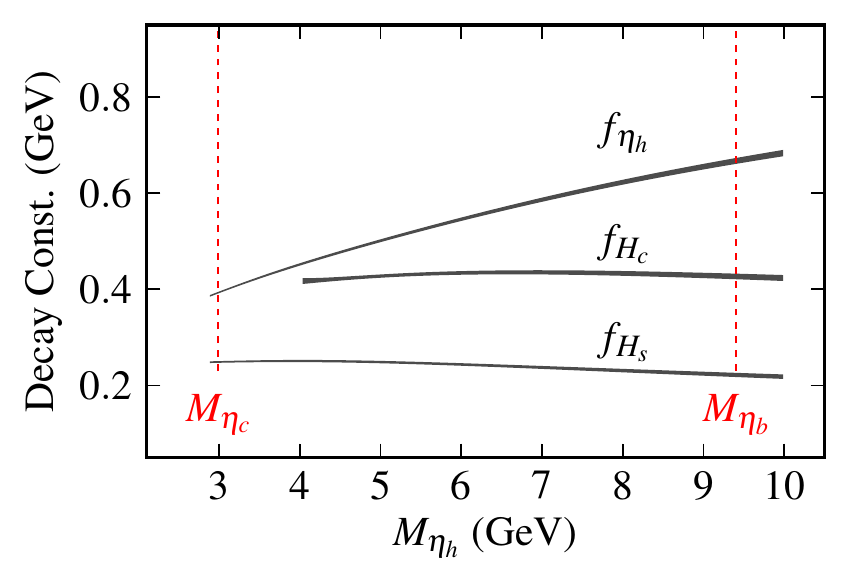}
\end{center}
\caption{ Summary of heavy quark mass dependence 
of decay constants for the pseudoscalar $H_c$, $H_s$ 
and $\eta_h$ mesons. The grey bands show our physical 
curves from Figures~\ref{fig:fhh},~\ref{fig:fhs} 
and~\ref{fig:fhc}. }
\label{fig:decayconstmh}
\end{figure}

\begin{figure*}
\begin{center}
\includegraphics[width=0.7\hsize]{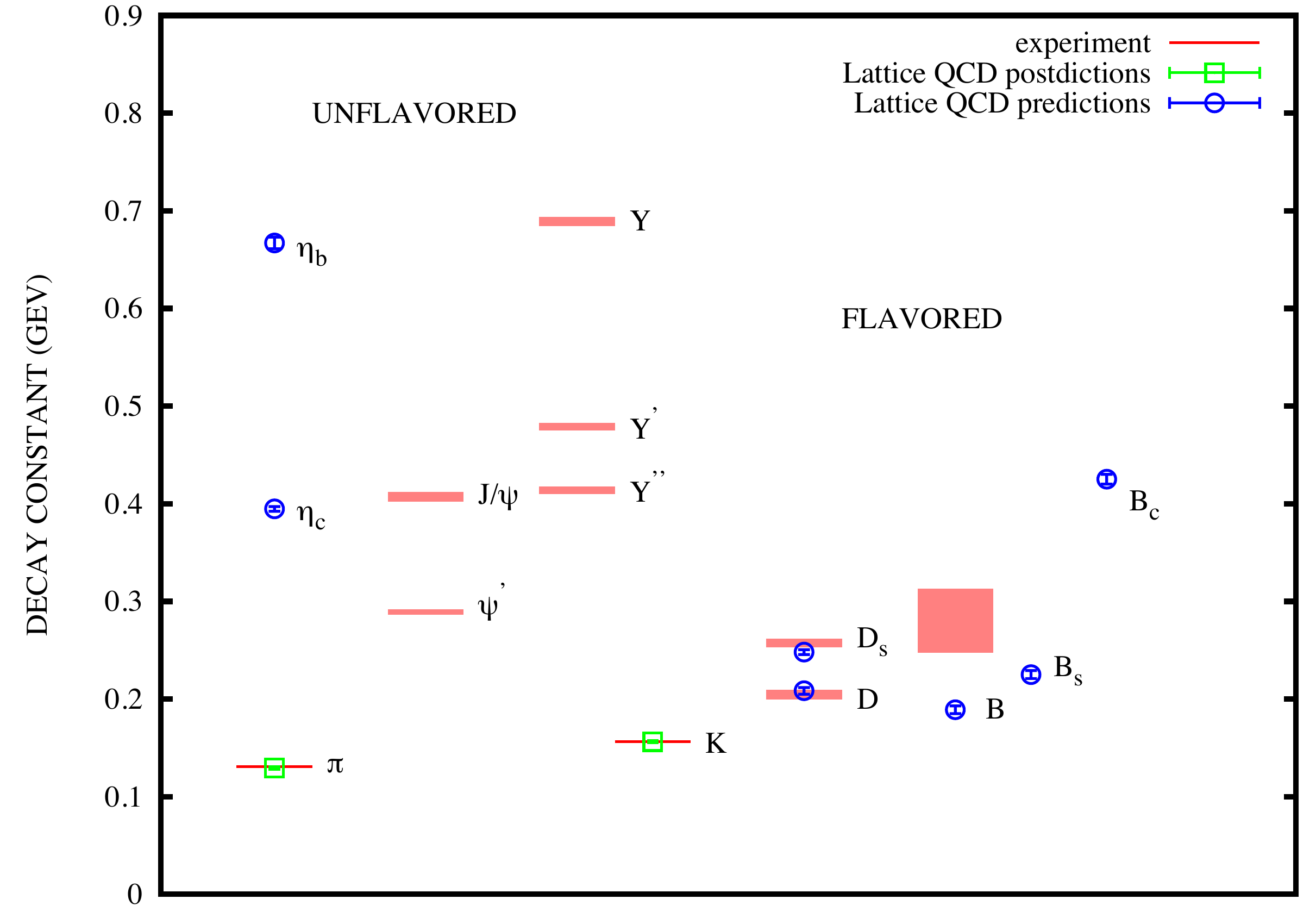}
\end{center}
\caption{ Spectrum of the decay constants of gold-plated particles 
from experiment (using values for 
CKM elements where needed) and from lattice QCD. 
Lattice QCD results are divided into 
postdictions (green open squares) and predictions (blue open 
circles). Results for $f_{B_c}$ and 
$f_{\eta_b}$ come from this paper, $f_{D_s}$ and $f_{\eta_c}$ 
from~\cite{fdsupdate}, $f_{\pi}$ and $f_K$ from~\cite{milc10}, 
$f_B$ from~\cite{newfb} and $f_D$ from~\cite{newfd}. 
Experimental results are given by red shaded bars. 
For unflavored vector mesons these come from~\cite{pdg} 
using eq.~(\ref{eq:vdecay}). $f_{\pi}$ and $f_K$ are from~\cite{pdg, rosnerstone}, 
$f_D$ is the updated experimental average from~\cite{bes3} 
and $f_{D_s}$, from~\cite{belle}. $f_B$ is not well-determined 
from experiment. We use the result from~\cite{cdlat11} using 
averages from~\cite{pdg}.  
}
\label{fig:summary}
\end{figure*}

By using a relativistic approach to heavy quarks (HISQ) 
which has relatively small discretisation errors we 
have been able to map out the dependence on heavy quark 
mass of the pseudoscalar heavyonium, heavy-strange and 
heavy-charm decay constants and the heavy-strange and 
heavy-charm meson masses, complementing 
results in~\cite{bcmasses, bshisq}.  

We find the heavyonium decay constant surprisingly close in 
value to the experimental results for the charmonium and 
bottomonium vector decay constants. Work is underway to 
confirm this result using NRQCD for the heavy quark and 
to establish accurate results for the corresponding 
vector decay constants in lattice QCD. Although the $\eta_h$ 
decay constant has no simple connection to an observed 
experimental rate, it is useful for comparison and calibration of 
lattice QCD calculations in heavy quark physics since it 
can be determined to 1\%, as we have done here. 

Our result for the $B_c$ meson mass agrees well using 
the two different mass splittings, $\Delta_{B_c,hh}$ and 
$\Delta_{B_c,hs}$ and also agrees with the experimental value. 
This is confirmation of our earlier result~\cite{gregory} using 
NRQCD $b$ quarks and HISQ light quarks. 

We determine the $B_c$ decay constant as 427(6) MeV, for the 
first time in full QCD, predicting
a leptonic branching ratio for the $B_c$ to $\tau \nu$ of 
1.9(2)\% (where the uncertainty comes from $t_{B_c}$, not $f_{B_c}$).
Our result for $f_{B_c}$ is significantly smaller 
than that from some potential model calculations, including 
those being used to estimate LHC production cross-sections~\cite{chang}. 
The best way to determine the $B_c$ leptonic 
decay rate, and hence $f_{B_c}$, from experiment 
may be using a high luminosity $e^+e^-$ collider 
operating at the $Z$ peak~\cite{akeroydzpeak, yangzpeak}.

By mapping out the dependence on the heavy quark mass of 
the $H_c$, $H_s$ and $\eta_h$ decay constants we 
are able to see the differences between the three systems. 
This is summarised in Figure~\ref{fig:decayconstmh} 
where we give the physical curves determined from 
our fits. In section~\ref{sec:discussion} we provide evidence 
that the $B_c$ behaves, at least in some ways, more like 
a heavy-light system than a heavy-heavy one. We previously 
noticed this effect in~\cite{ericbcstar} when 
finding that the mass difference between $B_c^*$ and $B_c$ 
was very close to the difference between $B_s^*$ and $B_s$. 

Table~\ref{tab:fitnums} gives results extracted from 
our fits at intermediate values of $M_{\eta_h}$ from 
$M_{\eta_c}$ to $M_{\eta_b}$ for comparison to future 
lattice QCD calculations or to phenomenological models. 
The values are determined by evaluating our fit function 
in the continuum limit and at tuned $s$ and $c$ masses, 
corresponding to the black line in 
Figs.~\ref{fig:fhh},~\ref{fig:mhs},~\ref{fig:fhs},~\ref{fig:deltabc},~\ref{fig:fhc} 
and~\ref{fig:mhcs}.

In Figure~\ref{fig:summary} we summarise the current picture 
for the decay constants of gold-plated mesons, determined 
from lattice QCD and from experiment. For lattice QCD we 
use the best existing results which dominate the world 
averages~\cite{cdlat11, fdsupdate, bshisq, newfb, newfd, milc10}. 
For the experimental values for the 
unflavored vectors we use leptonic widths to $e^+e^-$ from 
the Particle Data Tables~\cite{pdg} and eq.~(\ref{eq:vdecay}). 
For the flavored 
pseudoscalars the determination of the decay constant from 
experiment requires the input of a value for the associated 
CKM element, for example from the unitarity fit 
to the CKM matrix~\cite{pdg}. We update the $D$ and $D_s$
experimental determinations to the averages 
including new results from BESIII~\cite{bes3} and Belle~\cite{belle}
respectively. 

This plot goes beyond 
the traditional plot of the mass 
spectrum~\cite{cdlat11}  to look at a number 
which is related to the internal structure of the meson. 
The energy scale for decay constants is controlled by internal 
momenta inside the meson and so is much compressed over the 
scale for masses (which covers a large range simply because 
quark masses have a large range). The pseudoscalar meson 
decay constants are well filled in but more work is needed 
to obtain the vector decay constants to the same level of accuracy. 
This is underway and once complete, this plot will provide a 
very stringent test of QCD that would be impossible with 
any method other than lattice QCD.  

From our results here and in~\cite{bcmasses, bshisq} we see that 
the relativistic heavy quark approach using the HISQ formalism 
can successfully give results for the $b$ quark. Future work 
will use even finer lattices. For $a=$ 0.03fm, for example, the $b$ quark 
mass in lattice units is around 0.5 
and so we can easily achieve this mass without the need for 
extrapolation. 

{\bf{Acknowledgements}} We are grateful to the MILC collaboration 
for the use of 
their configurations and to C. Parkes for useful discussions. 
Computing was done at the Ohio Supercomputer Center, the Argonne 
Leadership Computing Facility at Argonne National Laboratory, supported
by the Office of Science of the U.S. Department of Energy under 
Contract DOE-AC02-06CH11357 and facilities of the DEISA consortium 
(www.deisa.eu) through the DEISA Extreme Computing 
Initiative, co-funded through the EU FP6 project RI-031513 and 
FP7 project RI-222919.
We acknowledge the use of Chroma~\cite{chroma} for part 
of our analysis. 
Funding for this work came from MICINN (grants FPA2009-09638 
and FPA2008-10732 and the Ramon y Cajal program), DGIID-DGA (grant 
2007-E24/2), the EU (ITN-STRONGnet, PITN-GA-2009-238353), the NSF, 
the Royal Society, the Wolfson Foundation, the Scottish Universities 
Physics Alliance and STFC. 

\bibliography{heavy}

\end{document}